\documentclass{aa}
\bibpunct{(}{)}{;}{a}{}{,} % to follow the A&A style
\usepackage{natbib}
\usepackage{amsmath}
\usepackage{physics}
\usepackage{graphicx}
\usepackage{float}
\usepackage{txfonts}
\begin{document}

    \title{The inverse Lidov-Kozai resonance \\ 
           for an outer test particle  due to an eccentric perturber}

   \author{G. C. de El\'\i a\inst{1,2}\thanks{gdeelia@fcaglp.unlp.edu.ar},
           M. Zanardi\inst{1,2},
           A. Dugaro\inst{1,2},
           \and
           S. Naoz\inst{3}
          }

   \offprints{G. C. de El\'\i a
    }

   \institute{Instituto de Astrof\'{\i}sica de La Plata, CCT La Plata-CONICET-UNLP \\
   Paseo del Bosque S/N (1900), La Plata, Argentina
   \and Facultad de Ciencias Astron\'omicas y Geof\'\i sicas, Universidad Nacional de La Plata \\
   Paseo del Bosque S/N (1900), La Plata, Argentina
   \and Department of Physics and Astronomy, University of California, Los Angeles, CA 90095, USA
                   }

   \date{Received / Accepted}
%-----------------------------------------------------------------------------------------------------------
%-----------------------------------------------------------------------------------------------------------

\abstract
%CONTEXT
{}
%AIMS
{We analyze the behavior of the argument of pericenter $\omega_2$ of an outer particle in the elliptical restricted three-body problem, focusing on the $\omega_2$ resonance or inverse Lidov-Kozai resonance.   
}
%METHODS
{First, we calculate the contribution of the terms of quadrupole, octupole, and hexadecapolar order of the secular approximation of the potential to the outer particle's $\omega_2$ precession rate $(d\omega_2/d\tau)$. Then, we derive analytical criteria that determine the vanishing of the $\omega_2$ quadrupole precession rate $(d\omega_2/d\tau)_{\text{quad}}$ for different values of the inner perturber's eccentricity $e_1$. Finally, we use such analytical considerations and describe the behavior of $\omega_2$ of outer particles extracted from N-body simulations developed in a previous work.
}
%RESULTS
{Our analytical study indicates that the values of the inclination $i_2$ and the ascending node longitude $\Omega_2$ associated with the outer particle that vanish $(d\omega_2/d\tau)_{\text{quad}}$ strongly depend on the eccentricity $e_1$ of the inner perturber. In fact, if $e_1 <$ 0.25 ($>$ 0.40825), $(d\omega_2/d\tau)_{\text{quad}}$ is only vanished for particles whose $\Omega_2$ circulates (librates). For $e_1$ between 0.25 and 0.40825, $(d\omega_2/d\tau)_{\text{quad}}$ can be vanished for any particle for a suitable selection of pairs ($\Omega_2$, $i_2$). Our analysis of the N-body simulations shows that the inverse Lidov-Kozai resonance is possible for small, moderate and high values of $e_1$. Moreover, such a resonance produces distinctive features in the evolution of a particle in the ($\Omega_2$, $i_2$) plane. In fact, if $\omega_2$ librates and $\Omega_2$ circulates, the extremes of $i_2$ at $\Omega_2 =$ 90$^{\circ}$ and 270$^{\circ}$ do not reach the same value, while if $\omega_2$ and $\Omega_2$ librate, the evolutionary trajectory of the particle in the ($\Omega_2$, $i_2$) plane evidences an asymmetry respect to $i_2 =$ 90$^{\circ}$. The evolution of $\omega_2$ associated with the outer particles of the N-body simulations can be very well explained by the analytical criteria derived in our investigation.
}
%CONCLUSIONS
{} 


\authorrunning{G. C. de El\'\i a, et al.
               }
\titlerunning{The inverse Lidov-Kozai resonance for an eccentric perturber}

\maketitle
%---------------------------------------------------------------------------------------------------------
%INTRODUCCION
%INTRODUCCION
\section{Introduction}

The approximation of secular dynamics is a powerful tool for studying a wide diversity of astrophysical problems. In particular, the works carried out by \citet{Lidov1962} and \citet{Kozai1962} were the first ones aimed at exploring the secular perturbations produced by a far away massive planet on a circular orbit over an inner test particle. In their works, the authors showed that, if the minimum inclination of the inner massless particle is between 39.23$\degr$ and 140.77$\degr$, its eccentricity $e$, inclination $i$, and argument of pericenter $\omega$ experience coupled oscillations. This secular effect associated with the evolution of the inner test particle is known as Lidov-Kozai mechanism.

When the hypothesis of the circular orbit for the outer perturber relaxes, interesting dynamical behaviors of the inner test particle become visible. Several authors have focused on the so-called inner eccentric Lidov-Kozai mechanism, analyzing the evolution of an inner particle in the elliptical restricted three-body problem. In particular, \citet{Harrington1968}, \citet{Soderhjelm1984}, \citet{Krymolowski1999}, \citet{Ford2000}, and  \citet{Naoz2013} derived general expressions for the secular hamiltonian, while \citet{Lithwick2011} and \citet{Katz2011} explored the case of the dynamical evolution of a inner test particle that orbits a star under the effects of an outer and eccentric massive perturber up to the octupole level of secular approximation. In such studies, it is possible understand that if the outer massive perturber has an eccentric orbit, the vertical angular momentum of the inner massless particle is not conserved and thus, its orbital plane can change the orientation from prograde to retrograde and back again reaching very high eccentricities close to the unity.  

The study of the secular dynamics of an outer test particle in the elliptical restricted three-body problem also shows significant results concerning the evolution of its orbital parameters. In this sense, \citet{Ziglin1975} was the first in to describe several aspects concerning the so-called outer eccentric Lidov-Kozai mechanism, analyzing the secular evolution of a distant planet orbiting a binary star system. From a doubly averaged disturbing function and under the context of the restricted three-body problem, the author showed that the evolution of the inclination $i$ and the ascending node longitude $\Omega$ of the outer planet is coupled, and furthermore, the width of libration region associated with $\Omega$ only depends on the eccentricity of the inner binary. Recently, \citet{Farago2010} analyzed the dynamical evolution for a distant test particle that orbits an central star under the gravitational effects of a internal massive planet up to the quadrupole level of the secular approximation. These authors confirmed the results obtained by \citet{Ziglin1975} and extended their studies to the general case of the three-body problem. 

The outer restricted three-body problem was also studied by \citet{LiD2014}, who obtained general solutions to some orbital parameters up to the quadrupole level of the secular approximation. Furthermore, \citet{LiD2014} explored the contribution of the octupole term of the secular hamiltonian and determined the evolution of the eccentricity of the outer massless particle for slightly inclined orbits and near polar configurations, and different values associated with the inner binary's eccentricity.

More recently, \citet{Naoz2017} and \citet{Zanardi2017} carried out two joint works aimed at studying the effects produced by an inner and eccentric massive perturber over the dynamical evolution of an outer massless particle, which orbit around a given central star. On the one hand, \citet{Naoz2017} derived analytical expressions up to the octupole level of the secular approximation for an outer test particle in the elliptical restricted three-body problem. From this, the authors showed that the analysis of the system under consideration is consistent with that derived by \citet{Ziglin1975} up to the quadrupole level of the approximation, and they discussed in detail the sensitivity of the octupole terms of the secular hamiltonian to the results. On the other hand, \citet{Zanardi2017} studied the formation and evolution of outer small body populations in systems that suffer strong scattering events between giant planets around 0.5 M$_{\odot}$ stars from N-body simulations. In particular, the authors focused the investigation on systems composed of a single inner Jupiter-mass planet and a far away reservoir of small bodies after the dynamical instability event. According to such an study, a natural result observed in these systems is the generation of particles with prograde and retrograde orbits whose ascending node longitude $\Omega$ circulates, and particles whose orbit plane flips\footnote{We define flipped orbits by those for which their inclination $i$ oscillates from below 90$^{\circ}$ to above 90$^{\circ}$. In other words, our definition focuses on the $i$ = 90$^{\circ}$ transition rather than the location and phase of $i$ in the surface of section.} from prograde to retrograde and back again throughout the evolution with coupled librations associated with $\Omega$. It is worth noting that the comparative analysis carried out by \citet{Zanardi2017} shows that the numerical results obtained in their research are supported by the analytical expressions derived by \citet{Naoz2017}. 

Then, \citet{Zanardi2018} analyzed the role of general relativity (GR) on outer small body reservoirs under the effects of an inner and eccentric Jupiter-mass planet around low-mass stars. From N-body simulations and analytical criteria, the authors showed that the GR may significantly modify the dynamical properties of the outer test particles. In fact, if the GR is included in the analysis, the range of prograde (retrograde) inclinations of the libration region associated with the ascending node longitude is reduced (increased) in comparison with that observed in absence of GR.   

Finally, \citet{Vinson2018} performed an study concerning the secular dynamics of an outer test particle in the elliptical restricted three-body problem from the expansion of the potential up to hexadecapolar level of the approximation. In such a research, the authors described the quadrupole orbital-flipping resonance, the octupole resonances associated with librations of $\omega + \Omega$ and $\omega-\Omega$ around 0$\degr$, and the hexadecapolar $\omega$ resonance, in which $\omega$ librates around 90$\degr$ or 270$\degr$, and it is so-called inverse Lidov-Kozai resonance.    

The inverse Lidov-Kozai resonance in the transneptunian region of the solar system was analyzed by \citet{Gallardo2012} assuming the gravitational effects of the giant planets on circular and coplanar orbits. Such as we mentioned in the previous paragraph, \citet{Vinson2018} extended the study of the inverse Lidov-Kozai resonance for the case of an inner perturber moving on an eccentric orbit. These authors found the inverse Lidov-Kozai resonance only up to a value of the inner perturber's eccentricity of 0.1 in N-body experiments. However, \citet{Zanardi2017} noted the existence of a few flipping-particles that experience simultaneous librations of the ascending node longitude $\Omega$ and the argument of pericenter $\omega$, which are associated with an inner perturber's eccentricity greater than 0.2 in their N-body simulations. 

From such results, the main goal of this research is to analyze in detail the inverse Lidov-Kozai resonance as a function of the inner perturber's eccentricity from analytical criteria and N-body simulations. Working on the basis of the elliptical restricted three-body problem, we will show that the inverse Lidov-Kozai resonance can be found for small, moderate, and high values of the orbital eccentricity associated with the inner perturber.   

The present paper is structured as follows. In Section 2, we present a brief overview up to date of the outer eccentric Lidov-Kozai mechanism. The analytical treatment concerning the precession rate of the argument of pericenter of an outer test particle is described in Section 3. A detailed analysis of results derived from N-body simulations is shown in Section 4. Finally, Section 5 describes the discussions and conclusions of our study.

%%%%%%%%%%%%%%%%%%%%%%%%%%%%%%%%%%%%%%%%%%%%%%%%%%%%%%%%%%%%%%%%%%%%%%%%%
%========================================================================
\section{Overview}

Recently, \citet{Naoz2017} analyzed in detail the secular dynamics of an outer test particle evolving under the gravitational influence of an inner and eccentric perturber of mass $m_1$ orbiting a central star of mass $m_0$. If the orbital parameters $e_2$, $i_2$, $\omega_2$, and $\Omega_2$ represent the eccentricity, the inclination with respect to the inner orbit, the argument of pericenter, and the ascending node longitude of the outer test particle relative to the inner perturber's periapse, respectively, the equations of motion can be written as partial derivatives of an energy function $f$ in the following way
\begin{eqnarray}
\frac{dJ_2}{d\tau} &=& \frac{\partial f}{\partial \omega_2} \\
\frac{dJ_{2,z}}{d\tau} &=& \frac{\partial f}{\partial \Omega_2} \\
\frac{d \omega_2}{d\tau} &=& \frac{\partial f}{\partial e_2}\frac{J_2}{e_2} + \frac{\partial f}{\partial \theta_2}\frac{\theta_2}{J_2} \\
\frac{d \Omega_2}{d\tau} &=& - \frac{\partial f}{\partial \theta_2}\frac{1}{J_2}
\end{eqnarray} 
where $\theta_2 = \cos i_2$, $J_2 = \sqrt{1-e^2_2}$, $J_{2,z} = \theta_2\sqrt{1-e^2_2}$, and $\tau$ is a parameter proportional to the true time $t$, which is given by $\tau = A t$, with 
\begin{eqnarray}
A = \frac{1}{16}\frac{m_0 m_1}{(m_0+m_1)^2}\sqrt{\frac{G(m_0+m_1)}{a^3_2}}\Bigg(\frac{a_1}{a_2}\Bigg)^2,
\label{factorA}
\end{eqnarray}
being $G$ the gravitational constant, and $a_1$ and $a_2$ the semimajor axis of the inner perturber and the outer test particle, respectively. The energy function $f$ derived by \citet{Naoz2017} up to the octupole level of the secular approximation is given by
\begin{equation}
f = f_{\text{quad}} + \epsilon f_{\text{oct}},
\label{fhastaoct}
\end{equation}
where the $\epsilon$ parameter is written by
\begin{equation}
\epsilon = \frac{m_0 - m_1}{m_0 + m_1}\frac{a_1}{a_2}\frac{e_2}{1-e^2_2},
\label{epsilon}
\end{equation}
and the functions $f_{\text{quad}}$ and $f_{\text{oct}}$\footnote{Note that \citet{Naoz2017} had a typo in their Eq. 7 concerning $f_{\text{oct}}$, which should have been preceded with a minus sign. It is worth noting that the equations of motion were calculated using the correct hamiltonian.} adopt the expressions
\begin{eqnarray}
f_{\text{quad}} &=& \frac{(2+3e^2_1)(3\theta^2_{2}-1)+15e^2_1(1-\theta^2_2)\cos(2\Omega_2)}{(1-e^2_2)^{3/2}} \\
f_{\text{oct}} &=& \frac{-15e_1}{4(1-e^2_2)^{3/2}}\Bigg\{10(1-e^2_1)\theta_2(1-\theta^2_2)\sin\omega_2\sin\Omega_2 \Bigg. \nonumber \\
             &+& \frac{1}{2}\bigg[2+19e^2_1-5(2+5e^2_1)\theta^2_2-35e^2_1(1-\theta^2_2)\cos(2\Omega_2)\bigg] \nonumber \\
             &\times& \Bigg.(\theta_2\sin\omega_2\sin\Omega_2-\cos\omega_2\cos\Omega_2)\Bigg\},
\end{eqnarray}
where $e_1$ is the eccentricity of the inner massive perturber. 

At the quadrupole-level of the secular approximation, \citet{Naoz2017} found that the inclination $i_2$ and the ascending node longitude $\Omega_2$ of the outer test particle evolve with precession rates given by $di_2/d\tau\propto\sin(2\Omega_2)$ and $d\Omega_2/d\tau\propto\theta_2$. According to this, on the one hand, the inclination $i_2$ adopts extreme values for $\Omega_2 =$ 0$^{\circ}$, 90$^{\circ}$ or 270$^{\circ}$. On the other hand, the ascending node longitude $\Omega_2$ reaches minimum and maximum values for $i_2 =$ 90$^{\circ}$, which represents a very important result to understand the secular behavior of an outer test particle. In fact, from this, the ascending node longitude $\Omega_2$ can evolve following two different regimes: circulation or libration. In a circulation trajectory, $\Omega_2$ adopts values between 0$^{\circ}$ and 360$^{\circ}$, and the outer test particle always evolves on prograde or retrograde orbits. In a libration trajectory, $\Omega_2$ is constrained between two specific values, and the orbital plane of the outer test particle flips from prograde to retrograde and back again along its evolution. This orbit-flipping quadrupole resonance, in which $\Omega_2$ librates around 90$^{\circ}$ or 270$^{\circ}$, was mentioned originally in the pioneer work carried out by \citet{Ziglin1975} and a detailed description about this can be found in \citet{Naoz2017}.   

The critical trajectory that divides the circulation and libration regimes in an inclination $i_2$ vs. ascending node longitude $\Omega_2$ plane is called separatrix. According to that described in last paragraph, the separatrix has associated a value of $i_2 =$ 90$^{\circ}$ for $\Omega_2 =$ 0$^{\circ}$. By taking into account that the extreme values of the inclination $i_2$ on the separatrix (hereafter referred by $i^{\text{e}}_2$) are obtained for $\Omega_2 =$ 90$^{\circ}$ or 270$^{\circ}$, the conservation of energy at the quadrupole-level of the approximation indicates that $f_{\text{quad}}(\Omega_2=0^{\circ},i_2=90^{\circ})=f_{\text{quad}}(\Omega_2=90^{\circ}$ or 270$^{\circ},i_2=i^{\text{e}}_2)$, where $f_{\text{quad}}$ is given by Eq. 8. From this, 
\begin{equation}
i^{\text{e}}_2 = \arccos\left\{\pm\sqrt{\frac{5e^2_1}{(1+4e^2_1)}}\right\},
\label{extremoi}
\end{equation}   
and then, the range of inclinations that lead to libration trajectories of $\Omega_2$ has a width $\Delta i_2$ given by 
\begin{equation}
\Delta i_2 = 2\arccos\left\{\sqrt{\frac{1-e^2_1}{(1+4e^2_1)}}\right\}.
\label{anchoi}
\end{equation}   
According to this expression, the more eccentric the inner perturber, the larger the range of inclinations that lead to libration trajectories of $\Omega_2$. In particular, it is worth remarking that a perturber on a circular orbit can not produce this orbit-flipping resonance that involves librations of $\Omega_2$ since the width $\Delta i_2 = 0^{\circ}$.

Recently, \citet{Vinson2018} studied the dynamics of an outer test particle in the secular three-body problem expanding the hamiltonian of the system up to hexadecapolar order. To do this, the authors adapted the disturbing function for an outer perturber derived by \citet{Yokoyama2003}. According to this, the hexadecapolar term of the secular approximation associated with the energy function $f$ for an outer test particle adopts the expression\footnote{It is important to remark that the terms of the energy function derived by \citet{Vinson2018} differ from those obtained by \citet{Naoz2017} in a constant factor. In particular, if we compare the quadrupole terms proposed in both works, it is possible to observe that the quadrupole term from \citet{Naoz2017} is obtained dividing the quadrupole term from \citet{Vinson2018} by $Gm_1a^2_1/(16a^3_2)$. Thus, we divide by such a factor the hexadecapolar term from \citet{Vinson2018} given by $c_3R_{\text{hex}}$ in order to use the equations of motion derived by \citet{Naoz2017} with the aim of calculating the hexadecapolar contribution of the secular approximation to the $\omega_2$ precession rate.}
\begin{equation}
f_{\text{hex}} = \frac{16}{G m_1}\frac{a^3_2}{a^2_1}c_3R_{\text{hex}}
\label{fhexa}
\end{equation}
where $c_3$ is given by
\begin{equation}
c_3 = G m_1\frac{a^4_1}{a^5_2}\frac{1}{(1-e^2_2)^{7/2}},
\label{eqc3}
\end{equation}
and 
\begin{eqnarray}
R_{\text{hex}} &=& \frac{3}{16}(2+3e^{2}_2)d_1-\frac{495}{1024}e^{2}_2-\frac{135}{256}\theta_2^2-\frac{165}{512} \nonumber \\
&+& \frac{315}{512}\theta_2^4+\frac{945}{1024}\theta_2^4e^{2}_2-\frac{405}{512}\theta_2^2e^{2}_2 \nonumber \\
&+& \Bigg\{\frac{105}{512}\theta_2^4+\frac{315}{1024}e^{2}_2-\frac{105}{256}\theta^2_2+\frac{105}{512}-\frac{315}{512}\theta_2^2e_2^2 \Bigg. \nonumber\\
&+& \Bigg.\frac{315}{1024}\theta^4_2e^2_2\Bigg\}d_3\cos(4\Omega_2) \nonumber \\
&+& \frac{105}{512}\Bigg\{\theta^3_2-\theta_2-\frac{1}{2}\theta^4_2+\frac{1}{2}\Bigg\}d_3e^2_2\cos(2\omega_2-4\Omega_2) \nonumber \\
&+& \frac{105}{512}\Bigg\{-\theta^3_2+\theta_2-\frac{1}{2}\theta^4_2+\frac{1}{2}\Bigg\}d_3e^2_2\cos(2\omega_2+4\Omega_2) \nonumber \\
&+& \Bigg\{\frac{45}{64}\theta^2_2-\frac{45}{512}-\frac{315}{512}\theta^4_2\Bigg\}e^2_2\cos(2\omega_2) \nonumber \\
&+& \Bigg\{\frac{15}{16}\theta^2_2+\frac{45}{32}\theta^2_2e^2_2-\frac{45}{256}e^2_2-\frac{15}{128}-\frac{315}{256}\theta^4_2e^2_2 \Bigg. \nonumber \\
&-& \Bigg.\frac{105}{128}\theta^4_2\Bigg\}d_2\cos(2\Omega_2) \nonumber \\
&+& \Bigg\{\frac{15}{256}-\frac{45}{128}\theta^2_2+\frac{75}{256}\theta_2+\frac{105}{256}\theta^4_2 \Bigg. \nonumber \\
&-& \Bigg.\frac{105}{256}\theta^3_2\Bigg\}d_2e^2_2\cos(2\omega_2-2\Omega_2) \nonumber \\
&+& \Bigg\{\frac{15}{256}-\frac{45}{128}\theta^2_2-\frac{75}{256}\theta_2+\frac{105}{256}\theta^4_2\Bigg. \nonumber \\
&+& \Bigg.\frac{105}{256}\theta^3_2\Bigg\}d_2e^2_2\cos(2\omega_2+2\Omega_2), 
\label{eqrex}
\end{eqnarray}
where
\begin{eqnarray}
d_1 &=& 1+\frac{15}{8}e^2_1+\frac{45}{64}e^4_1,  \\
d_2 &=& \frac{21}{8}e^2_1(2+e^2_1), \\
d_3 &=& \frac{63}{8}e^4_1. 
\end{eqnarray}
From the analysis carried out by \citet{Vinson2018}, it is worth noting that the $\omega_2$ resonance or inverse Lidov-Kozai resonance appears at hexadecapole order in the secular treatment due to the existence of the term in Eq.~\ref{eqrex} proportional to $e^2_2\cos(2\omega_2)$.

\citet{Vinson2018} derived interesting conclusions concerning the inverse Lidov-Kozai resonance for different values corresponding to the eccentricity $e_1$ of the inner perturber. In fact, from a secular analysis truncated up to the hexadecapolar order, the authors inferred that the inverse Lidov-Kozai resonance persists up to a value of $e_1 =$ 0.3. However, from N-body simulations, the authors found the inverse Lidov-Kozai resonance only up to a value of $e_1 =$ 0.1. From these results, \citet{Vinson2018} suggested that the inverse Lidov-Kozai resonance seems to disappear at higher $e_1$ being overwhelmed by the octupole effects. It is important to remark that the experiments carried out by these authors did not use a wide range of initial conditions associated with the orbital parameters of the the outer test particle for each value of the eccentricity of the inner perturber.

\section{Determination of the $\omega_2$ precession rate}

In the previous section, we describe the expression of the energy function $f$ associated with an outer test particle orbiting around a central star and evolving under the influence of an inner perturber with an arbitrary eccentricity. In particular, we present the quadrupole and octupole terms of the energy function calculated by \citet{Naoz2017}, and the hexadecapolar term derived by \citet{Vinson2018}. From such expressions and the equations of motion given by Eqs.~(1-4), it is possible to determine the change rate of the eccentricity $e_2$ ,the inclination $i_2$, the argument of pericenter $\omega_2$, and the ascending node longitude $\Omega_2$ associated with the outer test particle. As we have said before, our main research focuses on the inverse Lidov-Kozai resonance, for which we are only interested in computing the $\omega_2$ precession rate in the present study. 

\subsection{General treatment}

Taking into account the quadrupole, octupole, and hexadecapolar terms of the energy function, it is possible to determine the contribution of each order of the secular approximation to the $\omega_2$ precession rate. Thus, the $\omega_2$ quadrupole precession rate is given by
\begin{eqnarray}
\Bigg(\frac{d\omega_2}{d\tau}\Bigg)_{\text{quad}}\Bigg.&=&\frac{1}{(1-e^{2}_2)^{2}}\Bigg\{-6-9e^{2}_1+15\theta^{2}_2(2+3e^{2}_1)\Bigg.\nonumber\\ 
&+&\Bigg.15e^{2}_1(3-5\theta^{2}_2)\cos(2\Omega_2)\Bigg\}.
\label{tasaquad}
\end{eqnarray} 
Then, the contribution of the octupole term to the $\omega_2$ precession rate adopts the expression
\begin{eqnarray}
\Bigg(\frac{d\omega_2}{d\tau}\Bigg)_{\text{oct}}\Bigg.&=& \frac{\epsilon}{(1-e^2_2)^{1/2}}\Bigg\{(4+\frac{1}{e^2_2})f_{\text{oct}}\Bigg\} \nonumber \\
&-& \frac{15\epsilon e_1 \theta_2}{4(1-e^2_2)^2}\Bigg\{10(1-e^2_1)(1-3\theta^2_2)\sin\omega_2\sin\Omega_2 \nonumber \\
&+& \frac{1}{2}\bigg[-10\theta_2(2+5e^2_1)+70e^2_1\theta_2\cos(2\Omega_2)\bigg] \nonumber \\
&\times& (\theta_2\sin\omega_2\sin\Omega_2-\cos\omega_2\cos\Omega_2) \nonumber \\
&+& \frac{1}{2}\bigg[2+19e^2_1-5\theta^2_2(2+5e^2_1) \bigg.\nonumber \\ 
&-& \bigg.\Bigg.35e^2_1(1-\theta^2_2)\cos(2\Omega_2)\bigg]\sin\omega_2\sin\Omega_2\Bigg\}.
\label{tasaoct}
\end{eqnarray} 
Finally, the contribution of the hexadecapolar term to the $\omega_2$ precession rate is given by
\begin{eqnarray}
\Bigg(\frac{d\omega_2}{d\tau}\Bigg)_{\text{hex}}\Bigg.&=& \frac{16}{G m_1}\frac{a^3_2}{a^2_1}\frac{(1-e^2_2)^{1/2}}{e_2}\Bigg\{c_3 \frac{\partial R_{\text{hex}}}{\partial e_2}\Bigg. \nonumber \\ 
&+& \Bigg.\frac{e_2}{(1-e^2_2)}\theta_2c_3 \frac{\partial R_{\text{hex}}}{\partial \theta_2} 
+ R_{\text{hex}}\frac{\partial c_3}{\partial e_2} \Bigg\},
\label{tasahex}
\end{eqnarray} 
where $c_3$ and $R_{\text{hex}}$ are given by Eqs.~\ref{eqc3} and \ref{eqrex}, respectively, and 
\begin{eqnarray}
\frac{\partial R_{\text{hex}}}{\partial e_2} &=& \frac{3}{16}6e_2d_1-\frac{495}{1024}2e_2 + \frac{945}{1024}\theta_2^42e_2-\frac{405}{512}\theta_2^22e_2 \nonumber \\
&+& \Bigg\{\frac{315}{1024}2e_2-\frac{315}{512}\theta_2^22e_2 + \frac{315}{1024}\theta^4_22e_2\Bigg\}d_3\cos(4\Omega_2) \nonumber \\
&+& \frac{105}{512}\Bigg\{\theta^3_2-\theta_2-\frac{1}{2}\theta^4_2+\frac{1}{2}\Bigg\}d_32e_2\cos(2\omega_2-4\Omega_2) \nonumber \\
&+& \frac{105}{512}\Bigg\{-\theta^3_2+\theta_2-\frac{1}{2}\theta^4_2+\frac{1}{2}\Bigg\}d_32e_2\cos(2\omega_2+4\Omega_2) \nonumber \\
&+& \Bigg\{\frac{45}{64}\theta^2_2-\frac{45}{512}-\frac{315}{512}\theta^4_2\Bigg\}2e_2\cos(2\omega_2) \nonumber \\
&+& \Bigg\{\frac{45}{32}\theta^2_2-\frac{45}{256}-\frac{315}{256}\theta^4_2\Bigg\}d_22e_2\cos(2\Omega_2) \nonumber \\
&+& \Bigg\{\frac{15}{256}-\frac{45}{128}\theta^2_2+\frac{75}{256}\theta_2+\frac{105}{256}\theta^4_2-\frac{105}{256}\theta^3_2\Bigg\}\nonumber \\
&\times& d_22e_2\cos(2\omega_2-2\Omega_2) \nonumber \\
&+& \Bigg\{\frac{15}{256}-\frac{45}{128}\theta^2_2-\frac{75}{256}\theta_2+\frac{105}{256}\theta^4_2+\frac{105}{256}\theta^3_2\Bigg\}\nonumber \\
&\times& d_22e_2\cos(2\omega_2+2\Omega_2), \\
\frac{\partial R_{\text{hex}}}{\partial \theta_2} &=& -\frac{135}{256}2\theta_2+\frac{315}{512}4\theta_2^3+\frac{945}{1024}4\theta_2^3e^{2}_2-\frac{405}{512}2\theta_2e^{2}_2 \nonumber \\
&+& \Bigg\{\frac{105}{512}4\theta_2^3-\frac{105}{256}2\theta_2-\frac{315}{512}2\theta_2e_2^2 +\frac{315}{1024}4\theta^3_2e^2_2\Bigg\}\nonumber \\ &\times& d_3\cos(4\Omega_2) \nonumber \\
&+& \frac{105}{512}\Bigg\{3\theta^2_2-1-2\theta^3_2\Bigg\}d_3e^2_2\cos(2\omega_2-4\Omega_2) \nonumber \\
&+& \frac{105}{512}\Bigg\{-3\theta^2_2+1-2\theta^3_2\Bigg\}d_3e^2_2\cos(2\omega_2+4\Omega_2) \nonumber \\
&+& \Bigg\{\frac{45}{64}2\theta_2-\frac{315}{512}4\theta^3_2\Bigg\}e^2_2\cos(2\omega_2) \nonumber \\
&+& \Bigg\{\frac{15}{16}2\theta_2+\frac{45}{32}2\theta_2e^2_2-\frac{315}{256}4\theta^3_2e^2_2-\frac{105}{128}4\theta^3_2\Bigg\}\nonumber \\      
&\times& d_2\cos(2\Omega_2) \nonumber \\
&+& \Bigg\{-\frac{45}{128}2\theta_2+\frac{75}{256}+\frac{105}{256}4\theta^3_2-\frac{105}{256}3\theta^2_2\Bigg\}\nonumber \\
&\times& d_2e^2_2\cos(2\omega_2-2\Omega_2) \nonumber \\
&+& \Bigg\{-\frac{45}{128}2\theta_2-\frac{75}{256}+\frac{105}{256}4\theta^3_2+\frac{105}{256}3\theta^2_2\Bigg\}\nonumber \\ 
&\times& d_2e^2_2\cos(2\omega_2+2\Omega_2),  \\ 
\frac{\partial c_3}{\partial e_2} &=& 7 G m_1 e_2\frac{a^4_1}{a^5_2}\frac{1}{(1-e^2_2)^{9/2}}.
\end{eqnarray}
Once $(d\omega_2/d\tau)_{\text{quad}}$, $(d\omega_2/d\tau)_{\text{oct}}$, and $(d\omega_2/d\tau)_{\text{hex}}$ are calculated from Eqs.~\ref{tasaquad}, \ref{tasaoct}, and \ref{tasahex}, respectively, the $\omega_2$ precession rate up to the hexadecapolar order of the secular approximation is derived by
\begin{eqnarray}
\frac{d\omega_2}{d\tau} = \Bigg(\frac{d\omega_2}{d\tau}\Bigg)_{\text{quad}}+\Bigg(\frac{d\omega_2}{d\tau}\Bigg)_{\text{oct}}+\Bigg(\frac{d\omega_2}{d\tau}\Bigg)_{\text{hex}}.
\label{tasatotal}
\end{eqnarray}  
It is important to remark that if we want to compute the $\omega_2$ precession rate respect to the true time $t$, it is necessary to multiply Eq.~\ref{tasatotal} by the factor $A$ given by Eq.~\ref{factorA}.  

\subsection{Vanishing of the $\omega_2$ quadrupole precession rate}

In general terms, in a secular treatment, the quadrupole precession rate of the argument of pericenter of the outer test particle is that of greater magnitude since it is proportional to $(a_1/a_2)^2$. Thus, it is very interesting to determine the orbital parameter space that lead to the vanishing of $(d\omega_2/d\tau)_{\text{quad}}$. 

Setting $(d\omega_2/d\tau)_{\text{quad}} = 0$ in Eq.~\ref{tasaquad}, we find that the values of the inclination $i_2$ that satisfy this condition are given by
\begin{eqnarray}
i_2=\arccos\left\{\pm\sqrt{\frac{2+3e^2_1[1-5\cos(2\Omega_2)]}{10+5e^2_1[3-5\cos(2\Omega_2)]}}\right\}. 
\label{anulacionquadi}
\end{eqnarray} 
This expression allows us to derive several considerations of interest concerning the temporal evolution of $\omega_2$. The simplest case is produced when the orbit of the inner perturber is circular. In fact, if $e_1=0$, the quadrupole precession rate of the argument of pericenter of the outer test particle vanishes for orbital inclinations $i_2 =$ 63.4$^{\circ}$ and 116.6$^{\circ}$. This result is consistent with the critical inclinations that lead to the inverse Lidov-Kozai resonance derived by \citet{Gallardo2012}, who analyzed the Kozai dynamics in the transneptunian region of the solar system taking into account the gravitational perturbations of the giant planets moving in circular and coplanar orbits. More recently, \citet{Vinson2018} studied the outer elliptical restricted three-body problem and also determined such critical inclinations associated with the vanishing of the $\omega_2$ quadrupole precession rate analyzing the particular case of an inner perturber on a circular orbit.

\begin{figure*}
\centering
\includegraphics[angle=270, width=0.95\textwidth]{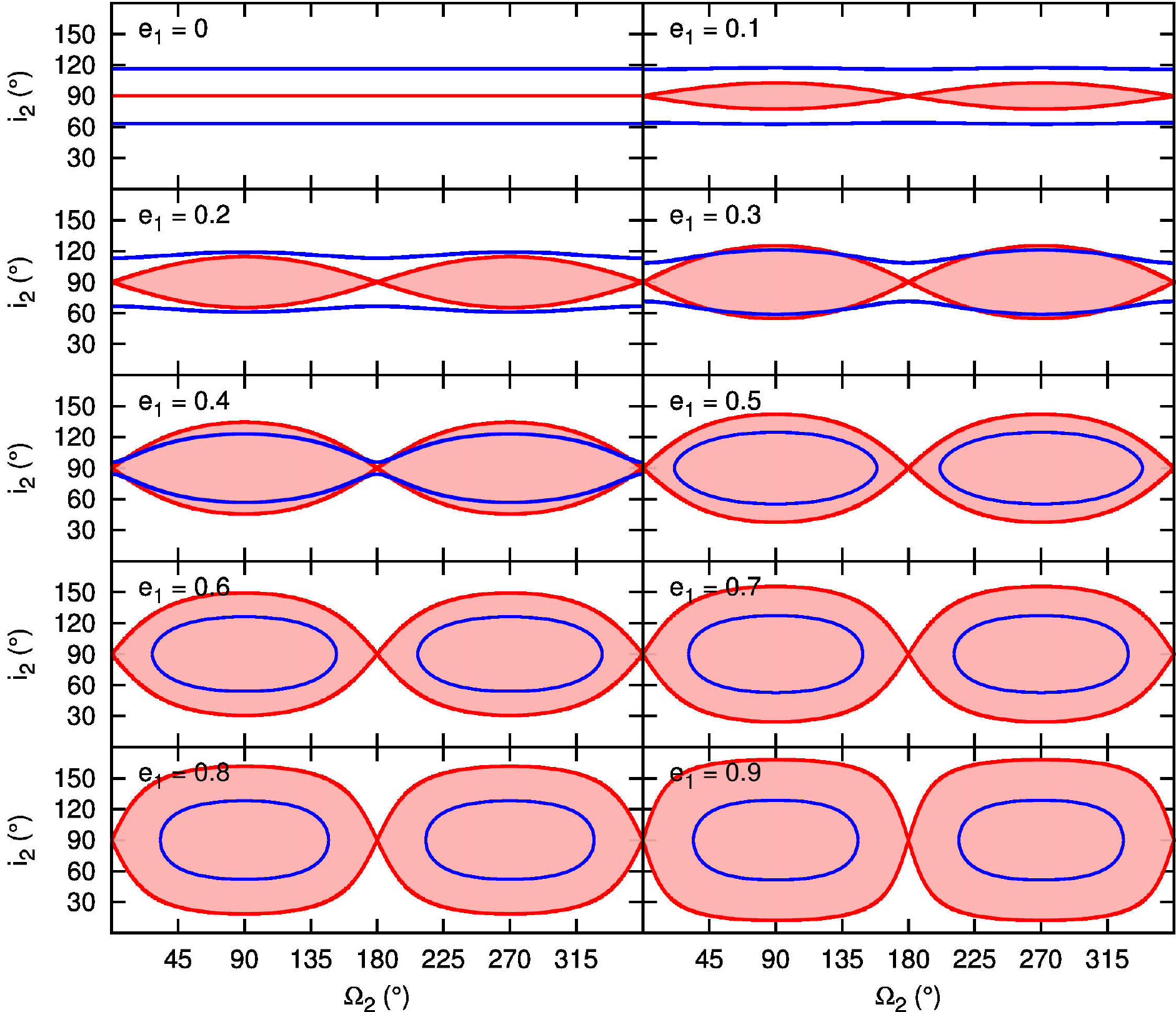}
\caption{
Pairs ($\Omega_2$, $i_2$) that vanish the $\omega_2$ quadrupole precession rate are illustrated by a blue curve for different values of the eccentricity $e_1$ associated with the inner perturber, which is specified at the top left corner of every panel. For each scenario, the separatrix of the system, which is computed up to the quadruple level of the secular approximation, is represented by a red curve, while the red shaded region illustrates the pairs ($\Omega_2$, $i_2$) that lead to the orbit-flipping quadrupole resonance.
}
\label{fig:fig1}
\end{figure*} 

The problem is more complex when the inner perturber's orbit is eccentric. In fact, if $e_1 > 0$, the values of the inclination $i_2$ that vanish the quadrupole precession rate of the argument of pericenter of the outer test particle depend on the ascending node longitude $\Omega_2$. The blue curve illustrated in Fig.~\ref{fig:fig1} represents the pairs ($\Omega_2$, $i_2$) that vanish the $\omega_2$ quadrupole precession rate for different values of the eccentricity $e_1$ associated with the inner perturber. Moreover, the red curve illustrates the separatrix, while the red shaded region represents the pairs ($\Omega_2$, $i_2$) that lead to the orbit-flipping resonance.

From this, Fig.~\ref{fig:fig1} allows us to appreciate important results concerning the vanishing of the $\omega_2$ quadrupole precession rate. On the one hand, for a value of $e_1 < e_{\text{1,NF}}$, all pairs ($\Omega_2$, $i_2$) that vanish the quadrupole precession rate of the argument of pericenter $\omega_2$ are located outside the red shaded region that corresponds to orbital flips. To calculate $e_{\text{1,NF}}$, we equate Eq.~\ref{anulacionquadi} at $\Omega_2 = 90^{\circ}$ and Eq.~\ref{extremoi}, which implies that $-64e^4_1 - 12e^2_1 + 1 = 0$. According to this expression, $e_{1,\text{NF}}$ adopts a value of 0.25.  

On the other hand, it is also important to remark that for a value of $e_1 > e_{1,\text{F}}$, the $\omega_2$ quadrupole precession rate is vanished only for pairs ($\Omega_2$, $i_2$) associated with the orbital flip red shaded region. Such as Fig.~\ref{fig:fig1} shows, to determine $e_{1,\text{F}}$, it is enough to evaluate the Eq.~\ref{anulacionquadi} at ($\Omega_2 = 0^{\circ}$, $i_2 = 90^{\circ}$), which is a pair corresponding to the separatrix. This condition implies that $2 - 12e^2_1 = 0$, which is satisfied for a value of $e_{1,\text{F}} =$ 0.40825. 

This simple analysis allows us to observe that the eccentricity $e_1$ of the inner perturber determines what kind of outer test particles vanish the $\omega_2$ quadrupole precession rate. In fact, if perturber's eccentricity $e_1 < e_{\text{1,NF}} = 0.25$, the $\omega_2$ quadrupole precession rate is only vanished for outer test particles on prograde and retrograde orbits whose $\Omega_2$ evolves on a circulatory regime, while when $e_1 > e_{\text{1,F}} = 0.40825$, only outer test particles on flipping orbits vanish the $\omega_2$ quadrupole precession rate. For perturbers with eccentricities $e_1$ between 0.25 and 0.40825, the $\omega_2$ quadrupole precession rate can be vanished for outer test particles on prograde and  retrograde orbits whose $\Omega_2$ circulates, as well as for particles that experience an orbit-flipping resonance, for values of the pair ($\Omega_2$, $i_2$) that satisfies the Eq.~\ref{anulacionquadi}.

In the next section, we make use of these analytical considerations with the aim of describing results obtained from N-body simulations. In fact, we use the expressions of the $\omega_2$ precession rate obtained in Sect. 3.1 and the criterion derived from the vanishing of the $\omega_2$ quadrupole precession rate to analyze the dynamical behavior of outer test particles that orbit around a central star and evolve under the effects of an inner and eccentric massive perturber in a given set of N-body experiments.      

%========================================================================
%         RESULTADOS
%========================================================================
\section{Results of N-body simulations}

\begin{figure*}
\centering
\includegraphics[width=0.95\textwidth]{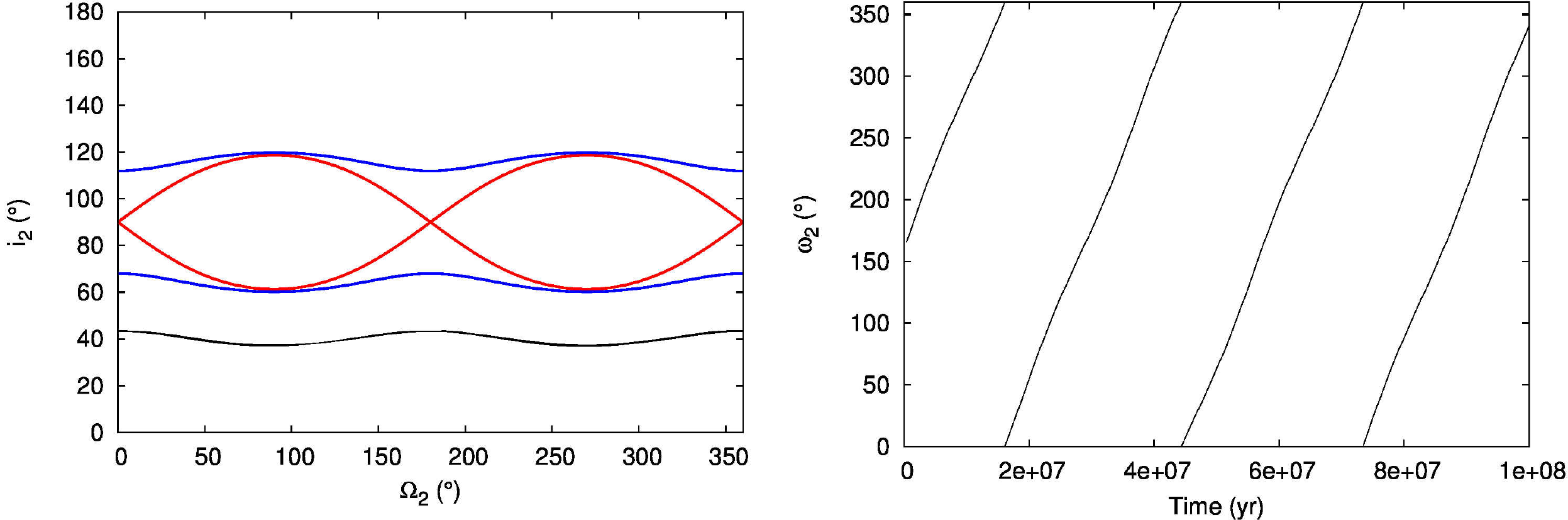} 
\caption{{\it Left panel:} Evolutionary trajectory of a Type-P particle associated with the Set 1 of N-body simulations is represented by the black curve in an inclination $i_2$ vs. ascending node longitude $\Omega_2$ plane. The initial orbital elements of this particle are $a_2$ = 21.935 au, $e_2$ = 0.452, $i_2$ = 37.129$^{\circ}$, $\omega_2$ = 165.966$^{\circ}$, and $\Omega_2$ = 97.285$^{\circ}$. The inner Jupiter-mass planet of such a system has a semimajor axis $a_1$ = 1.011 au and an eccentricity $e_1$ = 0.237. The red curve represents the separatrix of the system, while the blue curve illustrates the pairs ($\Omega_2$, $i_2$) that vanish the $\omega_2$ quadrupole precession rate for an inner perturber of $e_1$ = 0.237. {\it Right panel:} Temporal evolution of the argument of pericenter $\omega_2$ of the same particle.
}
\label{fig:fig59}
\end{figure*}

Recently, \citet{Zanardi2017} carried out N-body simulations\footnote{\citet{Zanardi2017} carried out the N-body simulations using the MERCURY code \citep{Chambers1999}. In particular, the authors used the RA15 version of the RADAU numerical integrator with an accuracy parameter of 10$^{-12}$ \citep{Everhart1985}.} aimed at studying the behavior of planetary systems that initially harbor three Jupiter-mass giants located close to their dynamical instability limit and an outer disk of test particles on quasi-circular and coplanar orbits around a 0.5 M$_{\odot}$ star. In this work, the authors analyzed the dynamical properties of such systems after undergoing strong planetary scattering events involving the three gaseous giants, and they focused their study on a set of N-body simulations, in which a single Jupiter-mass planet survives after the dynamical instability event. In particular, \citet{Zanardi2017} studied in detail a total of 12 N-body simulations, each of which began with 1000 test particles, being some of them removed during the evolution due to strong scattering events from collisions with the planets or the central star, or ejections from the system. In general terms, the 12 N-body simulations developed by \citet{Zanardi2017} produce systems with outer reservoirs composed of three different kinds of test particles: 1) particles on prograde orbits and whose $\Omega_2$ circulates (hereafter {\it Type-P particles}), 2) particles on retrograde orbits and whose $\Omega_2$ circulates (hereafter {\it Type-R particles}), and 3) particles whose orbital plane flips from prograde to retrograde and back again along their evolution and whose $\Omega_2$ librates (hereafter {\it Type-F particles}).

In the present work, we study the evolution of the argument of pericenter $\omega_2$ of the surviving outer test particles associated with the 12 N-body simulations analyzed in detail by \citet{Zanardi2017}, in each of which a single Jupiter-mass planet survives in the system with different values of the semimajor axis $a_1$ and the eccentricity $e_1$. In particular, our main research is based on outer test particles with values of $(a_1/a_2)$ and $\epsilon$ parameter $\lesssim$ 0.1. With these conditions, we seek to minimize the existence of non-secular effects in the sample of test particles of work extracted from the N-body simulations. 

It is important to remark that the Jupiter-mass planets surviving in the 12 N-body simulations that represent our sample of work show a wide diversity of orbital eccentricities. From such a sample, we can distinguish three different sets of N-body simulations: 
\begin{itemize}
\item {\it Set 1:} 2 of 12 N-body simulations, in which the inner Jupiter-mass planet has an eccentricity $e_1 < e_{\text{1,NF}} =$ 0.25, 
\item {\it Set 2:} 1 of 12 N-body simulations, in which the inner Jupiter-mass planet has an eccentricity 0.25 $< e_1 <$ 0.40825, and
\item {\it Set 3:} 9 of 12 N-body simulations, in which the inner Jupiter-mass planet has an eccentricity $e_1 > e_{\text{1,F}} =$ 0.40825,
\end{itemize}
Taking into account the analytical considerations derived in Sect. 3, we analyze the evolution of the argument of pericenter $\omega_2$ of the Type-P, -R, and -F outer test particles in each of these three different sets of N-body simulations. It is worth mentioning that we do not find test particle of interest concerning the evolution of the argument of pericenter $\omega_2$ in the N-body simulation associated with the Set 2. From this, our study focuses on the Sets 1 and 3 of N-body simulations, which show test particles with a wide diversity of dynamical behaviors.    

We want to remark two very important points concerning the N-body simulations used in the present research. On the one hand, the initial conditions of the test particles refer to orbital parameters immediately after the dynamical instability event, when a single Jupiter-mass planet survives in the system. In fact, we study the dynamical properties of each test particle in the frame of work of a restricted three body problem (star + planet + test particle) and its initial conditions must correspond to such a system. A detailed discussion about it is developed by \citet{Zanardi2017}, and an illustration of the initial conditions of the test particles associated with the 12 N-body simulations used in the present work can be found in Fig. 2 of \citet{Zanardi2018}. On the other hand, in order to compare properly the numerical results to the analytical expressions obtained in Sect. 3, we emphasize that the orbital parameters of the outer test particles and the surviving planet of each N-body simulation are assumed to be referenced to the barycenter and invariant plane of the system (star + planet), whose x-axis is selected to coincide with the surviving planet's periapse. To do this, we set the longitude of pericenter of the inner perturber $\varpi_1$ = 0 in agreement with \citet{Naoz2017} and \citet{Vinson2018}.

\subsection{Inner perturber with an eccentricity $e_1 < e_{\text{1,NF}} =$ 0.25}

\begin{figure}[!]
\centering
\includegraphics[width=0.48\textwidth]{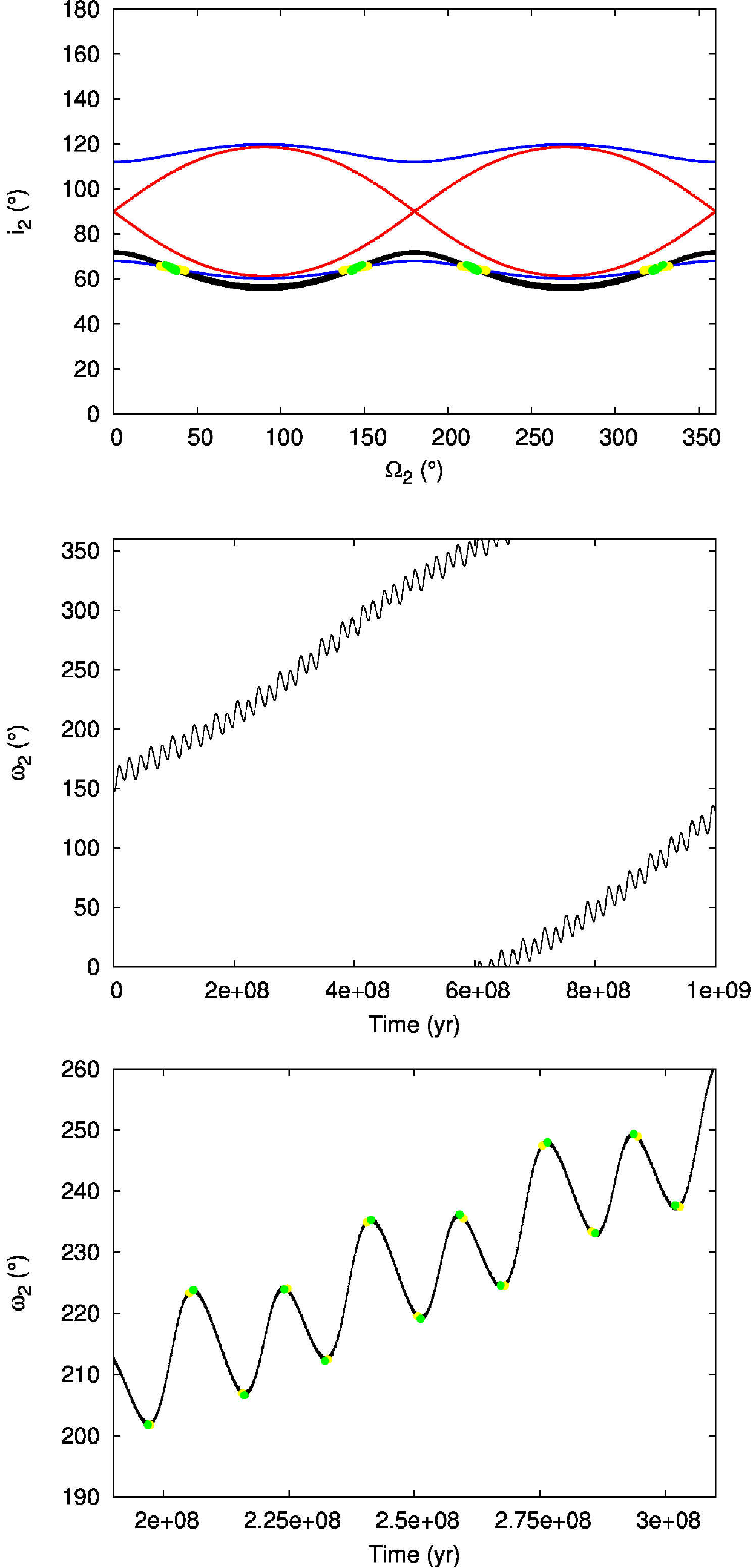} 
\caption{{\it Top panel:} Evolutionary trajectory of a Type-P particle associated with the Set 1 of N-body simulations in the ($\Omega_2$, $i_2$) plane (black curve). The initial orbital parameters of this particle are $a_2$ = 20.306 au, $e_2$ = 0.556, $i_2$ = 66.612$^{\circ}$, $\omega_2$ = 148.109$^{\circ}$, and $\Omega_2$ = 328.286$^{\circ}$. The values of $a_1$ and $e_1$ associated with the inner Jupiter-mass planet, and the red and blue curves are defined in the caption of Fig.~\ref{fig:fig59}. The yellow and green circles illustrate the pairs ($\Omega_2$, $i_2$) of the particle's trajectory associated with the vanishing of $(d\omega_2/d\tau)_{\text{quad}}$ and $(d\omega_{2}/d\tau)$, respectively. {\it Middle panel:} Evolution in time of the argument of pericenter $\omega_2$ of such a particle. {\it Bottom panel:} Zoom of the middle panel between 200 Myr and 300 Myr. The yellow and green circles represent the values of $\omega_2$ of the particle's trajectory associated with the vanishing of $(d\omega_2/d\tau)_{\text{quad}}$ and $(d\omega_{2}/d\tau)$, respectively.    
}
\label{fig:fig19}
\end{figure}

As we said above, in 2 of 12 N-body simulations, the eccentricity $e_1$ of the inner Jupiter-mass planet is less than $e_{\text{1,NF}} =$ 0.25. Specifically, the values of $e_1$ associated with such simulations are of 0.227 and 0.237. We carry out a detailed analysis of the outer reservoirs of such systems with the aim of describing the evolution of the argument of pericenter $\omega_2$ of the the Type-P, -R and -F particles.

\begin{figure}[!]
\centering
\includegraphics[width=0.48\textwidth]{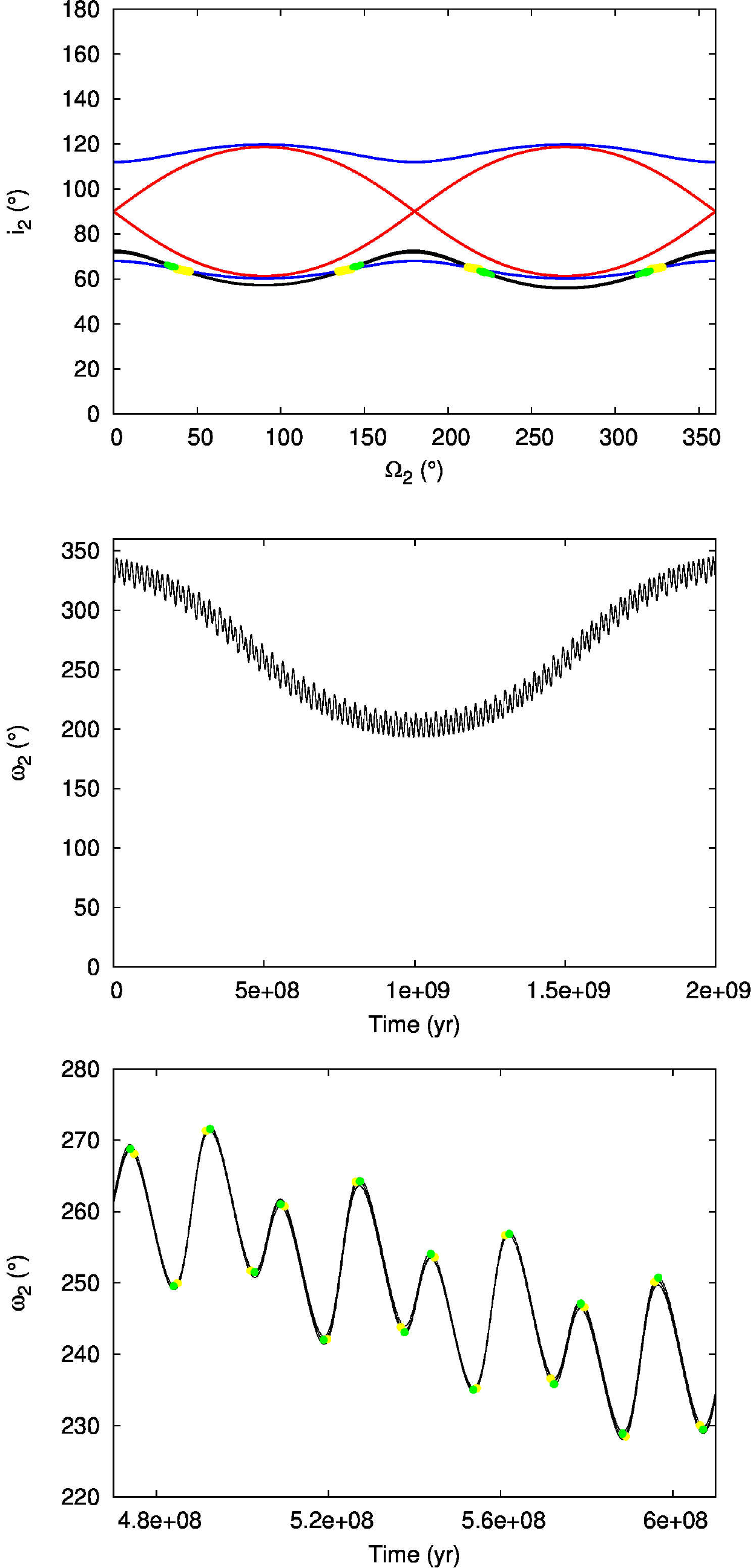} 
\caption{{\it Top panel:} Trajectory of a Type-P particle associated with the Set 1 of N-body simulations in the ($\Omega_2$, $i_2$) plane (black curve). The initial orbital parameters of this particle are $a_2$ = 17.424 au, $e_2$ = 0.371, $i_2$ = 69.930$^{\circ}$, $\omega_2$ = 325.686$^{\circ}$, and $\Omega_2$ = 159.785$^{\circ}$. The values of $a_1$ and $e_1$ associated with the inner perturber, and the red and blue curves are defined in the caption of Fig.~\ref{fig:fig59}. {\it Middle panel:} Temporal evolution of the argument of pericenter $\omega_2$ of such a particle, which librates by defining an inverse Lidov-Kozai resonance. {\it Bottom panel:} Zoom of the middle panel between 480 Myr and 600 Myr. The yellow and green circles illustrated in the top and bottom panels are defined in the caption of Fig.~\ref{fig:fig19}. 
}
\label{fig:fig778}
\end{figure}

In particular, Figs.~\ref{fig:fig59}, \ref{fig:fig19}, and \ref{fig:fig778} show the evolution of three different kinds of Type-P particles associated with a simulation whose inner perturber has an eccentricity $e_1$ of 0.237. The left panel of Fig.~\ref{fig:fig59} describes the evolution of a given Type-P particle in the ($\Omega_2$, $i_2$) plane, which is represented by the black curve. Moreover, the red curve illustrates the separatrix of the system, which is computed up to the quadrupole level of the secular approximation such as it was described in Sect. 2, and the blue curve shows the pairs ($\Omega_2$, $i_2$) that vanish the $\omega_2$ quadrupole precession rate for an inner perturber with an eccentricity $e_1$ of 0.237. As the reader can see, the ascending node longitude $\Omega_2$ of the test particle evolves in a circulation mode, while its inclination $i_2$ always adopts prograde values. It is important to remark that the trajectory of the test particle never crosses the blue curve, which indicates that its $\omega_2$ quadrupole precession rate does not vanish along its evolution. From this, the right panel of Fig.~\ref{fig:fig59} shows that the argument of pericenter $\omega_2$ of the test particle circulates over 10$^{8}$ yr of evolution.         

In the same way, the black curve in the top panel of Fig.~\ref{fig:fig19} illustrates the evolution of a test particle representative of a second kind of Type-P particles in the inclination $i_2$ vs. ascending node longitude $\Omega_2$ plane. In this case, the evolutionary trajectory of the particle periodically crosses the blue curve, which determines pairs ($\Omega_2$, $i_2$) that lead to the vanishing of the $\omega_2$ quadrupole precession rate. The middle panel of Fig.~\ref{fig:fig19} shows that the argument of pericenter $\omega_2$ of this Type-P particle circulates reaching local minimum and maximum values. 

To understand the existence of these local minimum and maximum associated with the evolution of argument of pericenter $\omega_2$ of the outer test particle under consideration, we compute the values of $i_2$, $\Omega_2$, and $\omega_2$ corresponding to the vanishing of the $\omega_2$ quadrupole precession rate $(d\omega_2/d\tau)_{\text{quad}}$, which is given by Eq.~\ref{tasaquad}, and the vanishing of the $\omega_2$ precession rate calculated up to the hexadecapolar level of the secular approximation $(d\omega_2/d\tau)$, which is represented by Eq.~\ref{tasatotal}. From this, on the one hand, the bottom panel of Fig.~\ref{fig:fig19} shows a zoom of the middle panel between 200 Myr and 300 Myr and it illustrates the values of $\omega_2$ associated with the vanishing of $(d\omega_2/d\tau)_{\text{quad}}$ and $(d\omega_{2}/d\tau)$ by yellow and green circles, respectively. As the reader can see, the local minimum and maximum observed in the evolution of $\omega_2$ are in a good agreement with the green circles, while the yellow ones are very close to them. On the other hand, the top panel of Fig.~\ref{fig:fig19} shows the pairs ($\Omega_2$, $i_2$) associated with the vanishing of $(d\omega_2/d\tau)_{\text{quad}}$ and $(d\omega_2/d\tau)$ by yellow and green circles, respectively. From this, the green circles represent values of $i_2$ and $\Omega_2$ very similar to those illustrated by the yellow ones, which are located on the blue curve.   

\begin{figure}[!]
\centering
\includegraphics[width=0.48\textwidth]{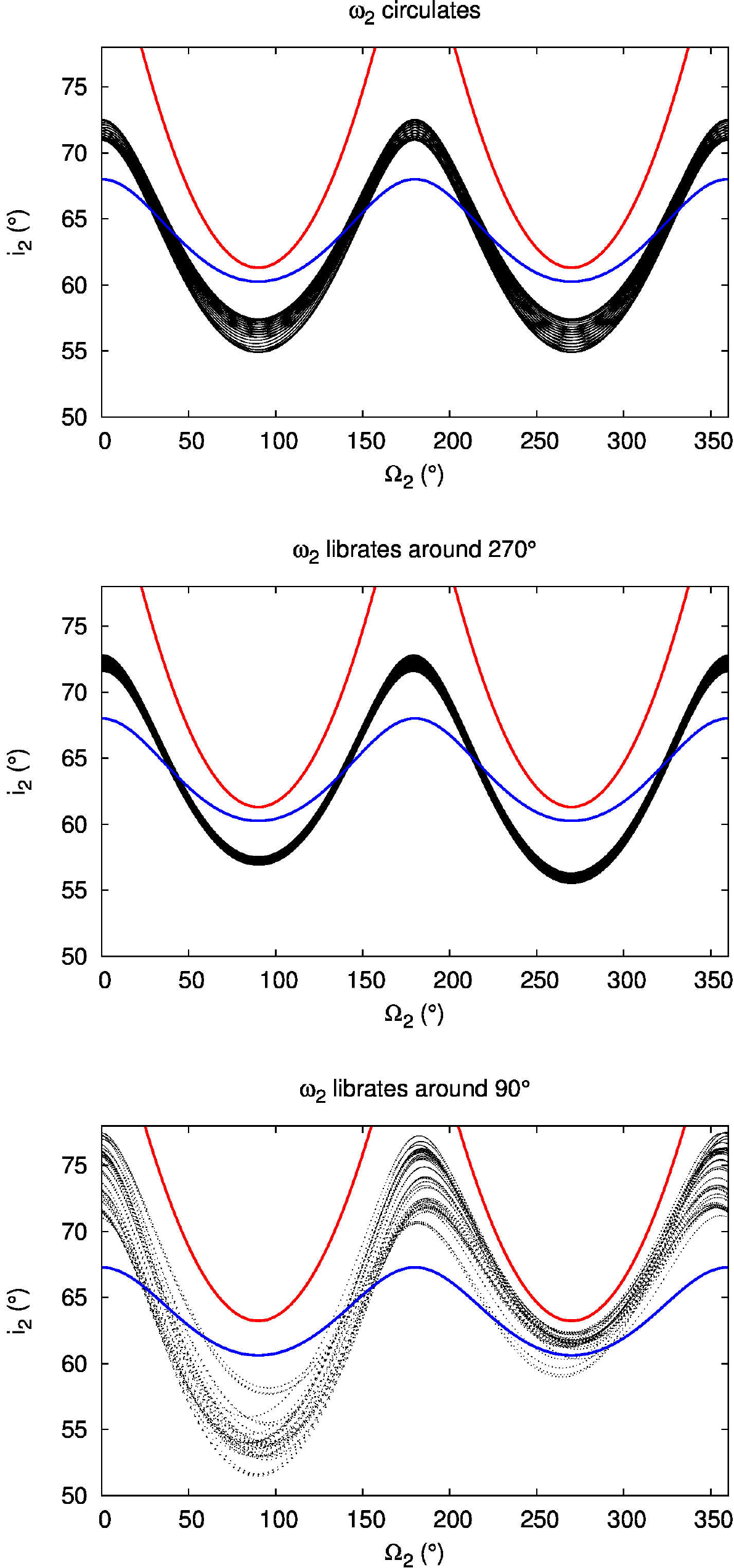} 
\caption{
Evolutionary trajectories in the ($\Omega_2$, $i_2$) plane of Type-P particles associated with the Set 1 of N-body simulations are displayed in a zoom. In every panel, the black curve illustrates the test particle's trajectory, while the red and blue curves represent the separatrix and the pairs ($\Omega_2$, $i_2$) that vanish the $\omega_2$ quadrupole precession rate of the system under consideration. The particles illustrated in the top and middle panels are those represented in Figs.~\ref{fig:fig19} and \ref{fig:fig778}, respectively. For the case shown in the bottom panel, the initial orbital parameters of the test particle are $a_2$ = 13.833 au, $e_2$ = 0.672, $i_2$ = 63.712$^{\circ}$, $\omega_2$ = 33.2$^{\circ}$, and $\Omega_2$ = 138.95$^{\circ}$, while the values of $a_1$ and $e_1$ of the inner Jupiter-mass planet are 1.561 au and 0.227, respectively. 
}
\label{fig:figcomparacion_incli_directas}
\end{figure}  

Finally, the top panel of Fig.~\ref{fig:fig778} shows the evolution of a test particle representative of a third kind of Type-P particles in the ($\Omega_2$, $i_2$) plane, which is illustrated by the black curve. According to this, the intersection points of the evolutionary trajectory of the particle with the blue curve define pairs ($\Omega_2$, $i_2$) that vanish the $\omega_2$ quadrupole precession rate. Once this is done, we analyze the temporal evolution of the argument of pericenter $\omega_2$ of the test particle, which is illustrated in the middle panel of Fig.~\ref{fig:fig778}. In this particular case, $\omega_2$ librates around 270$^{\circ}$, reaching local minimum and maximum values. This behavior defines an inverse Lidov-Kozai resonance, and it represents the only example that we find in the sample of particles with values of $(a_1/a_2)$ and $\epsilon$ parameter less than 0.1 in our 2 N-body simulations with an inner perturber whose eccentricity $e_1$ is less than $e_{\text{1,NF}} =$ 0.25. In fact, the inverse Lidov Kozai resonance is more restrictive than the conventional Lidov-Kozai resonance, and the orbital parameters (in particular, the values of $\Omega_2$ and $i_2$) that lead to an $\omega_2$ resonance for an outer test particle are more limited.             

For this particular case, we also derive the values of the inclination $i_2$, the ascending node longitude $\Omega_2$, and the argument of pericenter $\omega_2$ associated with the vanishing of the $\omega_2$ quadrupole precession rate $(d\omega_2/d\tau)_{\text{quad}}$ (yellow circles), and the $\omega_2$ precession rate computed up to the hexadecapolar level of the secular approximation $(d\omega_2/d\tau)$ (green circles). On the one hand, the yellow circles illustrated in the top panel of Fig.~\ref{fig:fig778} are logically located on the blue curve, while the green ones represent pairs ($\Omega_2$, $i_2$) close to them. On the other hand, the bottom panel of Fig.~\ref{fig:fig778} shows that the local minimum and maximum values of the temporal evolution of $\omega_2$ are in a good agreement with the green circles, but the yellow ones are close. 

We must remark a very important difference observed between Type-P particles whose argument of pericenter $\omega_2$ circulates, and those Type-P particles that experience an inverse Lidov-Kozai resonance. In fact, on the one hand, if $\omega_2$ circulates, the minimum values of the orbital inclination $i_2$ at $\Omega_2 =$ 90$^{\circ}$ and 270$^{\circ}$ are the same. This result is very well illustrated in the left and top panels of Figs.~\ref{fig:fig59} and \ref{fig:fig19}, respectively. Beyond this, the top panel of Fig.~\ref{fig:figcomparacion_incli_directas}, which represents a zoom of the top panel of Fig.~\ref{fig:fig19}, allows us a better appreciation of the equivalence between the minimum values of the inclination $i_2$ at $\Omega_2 =$ 90$^{\circ}$ and 270$^{\circ}$ of a test particle whose $\omega_2$ evolves in a circulatory regime. On the other hand, if $\omega_2$ librates, the minimum values of the orbital inclination $i_2$ at $\Omega_2 =$ 90$^{\circ}$ and 270$^{\circ}$ are different, which can be seen in the middle and bottom panels of Fig.~\ref{fig:figcomparacion_incli_directas}, which illustrate evolutionary trajectories of two different particles whose $\omega_2$ librates around 270$^{\circ}$ and 90$^{\circ}$, respectively. According to that observed in such panels, the value of the ascending node longitude $\Omega_2$ associated with the absolute minimum of the inclination $i_2$ of a Type-P particle in a $\omega_2$ resonance depends on the libration center of $\omega_2$. In fact, if an outer test particle experiences an inverse Lidov-Kozai resonance, the minimum value of the inclination $i_2$ corresponds to $\Omega_2 =$ 270$^{\circ}$ (90$^{\circ}$), if $\omega_2$ librates around 270$^{\circ}$ (90$^{\circ}$). We would like to remark that the middle panel of Fig.~\ref{fig:figcomparacion_incli_directas} is a zoom of the top panel of Fig.~\ref{fig:fig778}, which illustrates the evolution of the only particle with values of $(a_1/a_2)$ and $\epsilon$ parameter less than 0.1 in an inverse Lidov-Kozai resonance in our 2 N-body simulations with an inner perturber's eccentricity $e_1 < e_{\text{1,NF}} =$ 0.25. The bottom panel of Fig.~\ref{fig:figcomparacion_incli_directas} is associated with the evolution of a test particle that experiences a $\omega_2$ resonance with values of $(a_1/a_2)$ and $\epsilon$ parameter slightly greater than 0.1 in a system whose inner perturber's eccentricity $e_1 =$ 0.227. While our main investigation does not take into account such test particles, we decide to include it in order to explicitly show the sensitivity of the results exposed in Fig.~\ref{fig:figcomparacion_incli_directas} to the libration center of $\omega_2$.    

\begin{figure*}[!]
\centering
\includegraphics[width=0.95\textwidth]{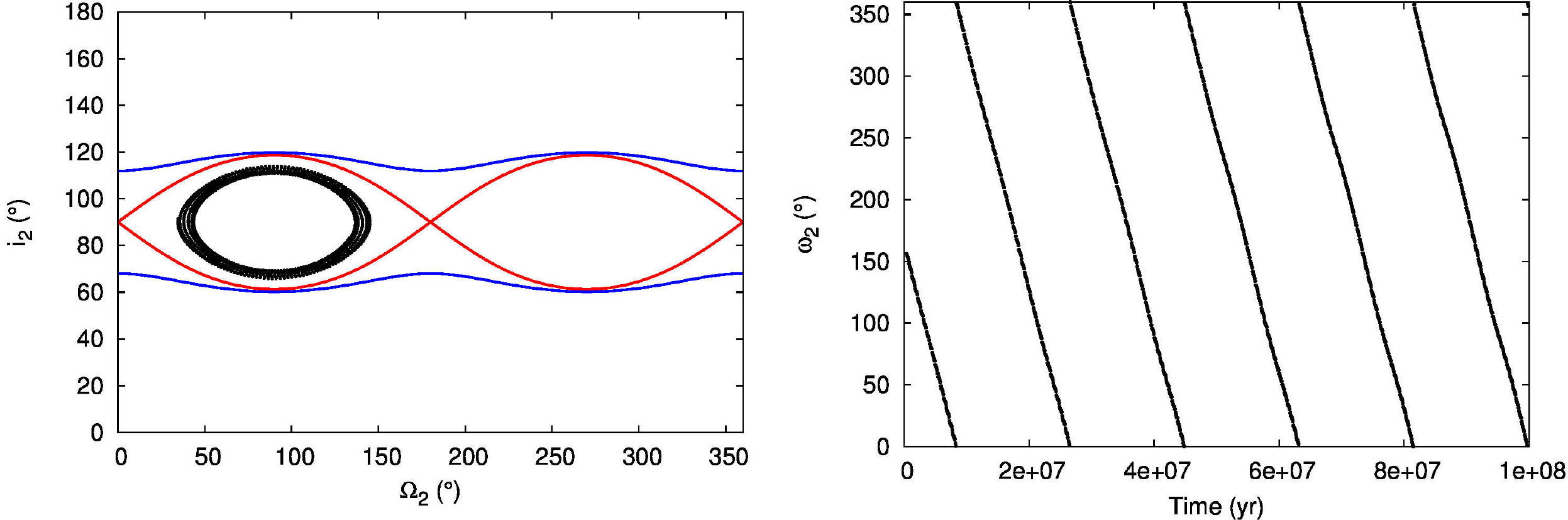} 
\caption{
{\it Left panel:} Trajectory of a Type-F particle corresponding to the Set 1 of N-body simulations in the ($\Omega_2$, $i_2$) plane (black curve). The initial orbital elements of this particle are $a_2$ = 14.116 au, $e_2$ = 0.262, $i_2$ = 83.136$^{\circ}$, $\omega_2$ = 156.0$^{\circ}$, and $\Omega_2$ = 133.263$^{\circ}$. The values of $a_1$ and $e_1$ associated with the inner perturber, and the red and blue curves are defined in the caption of Fig.~\ref{fig:fig59}. {\it Right panel:} Evolution in time of the argument of pericenter $\omega_2$ of the particle.  
}
\label{fig:fig905}
\end{figure*}

The Type-R outer test particles observed in this set of N-body simulations show a behavior similar to that described for the Type-P particles illustrated in Figs.~\ref{fig:fig59} and \ref{fig:fig19} concerning the evolution of the argument of pericenter $\omega_2$. In fact, as we said above, we do not find Type-R outer test particles that experience an inverse Lidov-Kozai resonance in our sample of work. It is important to remembering that the initial conditions of the test particles of the N-body simulations used in the present study correspond to orbital parameters immediately after the dynamical instability event, when a single Jupiter-mass planet survives in the system. Such as \citet{Zanardi2017} described and \citet{Zanardi2018} illustrated in their Fig. 2, the number of particles with initial orbital inclinations higher than 90$^{\circ}$ is much lower than the number associated with initial inclinations less than 90$^{\circ}$. According to this and taking into account the restrictive conditions associated with an inverse Lidov-Kozai resonance, it is easy to understand that none Type-R test particle was found in a $\omega_2$ resonance in the present set of N-body simulations.

Finally, in general terms, the behavior of the argument of pericenter $\omega_2$ of the Type-F particles associated with the Set 1 of N-body simulations is very simple. In fact, Fig.~\ref{fig:fig905} shows the evolution of a given Type-F particle associated with a system with an inner perturber whose eccentricity $e_1$ is 0.237. The left panel illustrates the trajectory of such a particle in an inclination $i_2$ vs. ascending node longitude $\Omega_2$ plane. According to \citet{Naoz2017} and \citet{Zanardi2017}, if the inclination $i_2$ of the outer particle's orbital plane flips from prograde to retrograde values and back again, the ascending node longitude $\Omega_2$ librates between two specific values. Thus, if the orbital plane of the particle flips along the entire evolution, its evolutionary trajectory is confined to the inner region to the separatrix in an $i_2$ vs. $\Omega_2$ plane, and it never crosses the blue curve, which indicates that the $\omega_2$ quadrupole precession rate does not vanish. From this, we analyze the temporal evolution of the argument of pericenter $\omega_2$ of such a Type-F particle. Our study shows that $\omega_2$ circulates over 10$^{8}$ yr, which is illustrated in the right panel of Fig.~\ref{fig:fig905}.    

\subsection{Inner perturber with an eccentricity $e_1 > e_{\text{1,F}} =$ 0.40825}

\begin{figure*}
\centering
\includegraphics[width=0.95\textwidth]{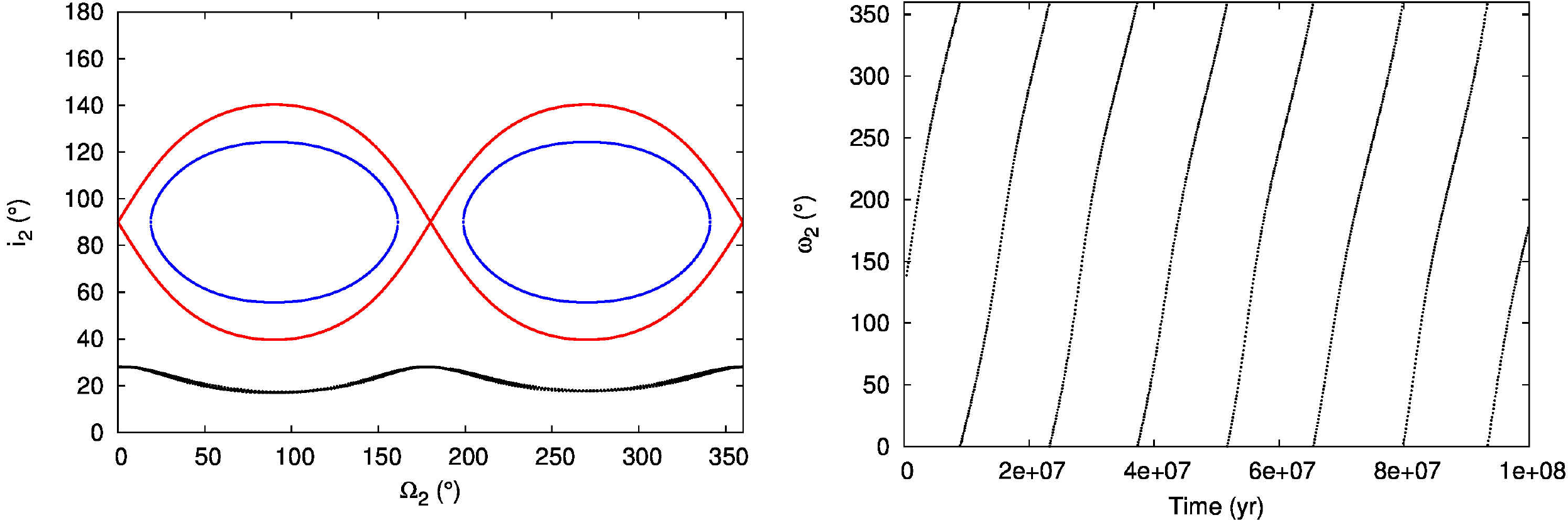} 
\caption{
{\it Left panel:} Trajectory of a Type-P particle associated with the Set 3 of N-body simulations in the ($\Omega_2$, $i_2$) plane (black curve). The initial orbital elements of this particle are $a_2$ = 38.098 au, $e_2$ = 0.676, $i_2$ = 18.455$^{\circ}$, $\omega_2$ = 134.168$^{\circ}$, and $\Omega_2$ = 119.149$^{\circ}$. The inner Jupiter-mass planet of such a system has a semimajor axis $a_1$ = 1.778 au and an eccentricity $e_1$ = 0.475. The red curve represents the separatrix of the system, while the blue curve illustrates the pairs ($\Omega_2$, $i_2$) that vanish the $\omega_2$ quadrupole precession rate for an inner perturber of $e_1$ = 0.475. {\it Right panel:} Temporal evolution of the argument of pericenter $\omega_2$ of the particle.  
}
\label{fig:fig298}
\end{figure*}

\begin{figure*}
\centering
\includegraphics[width=0.95\textwidth]{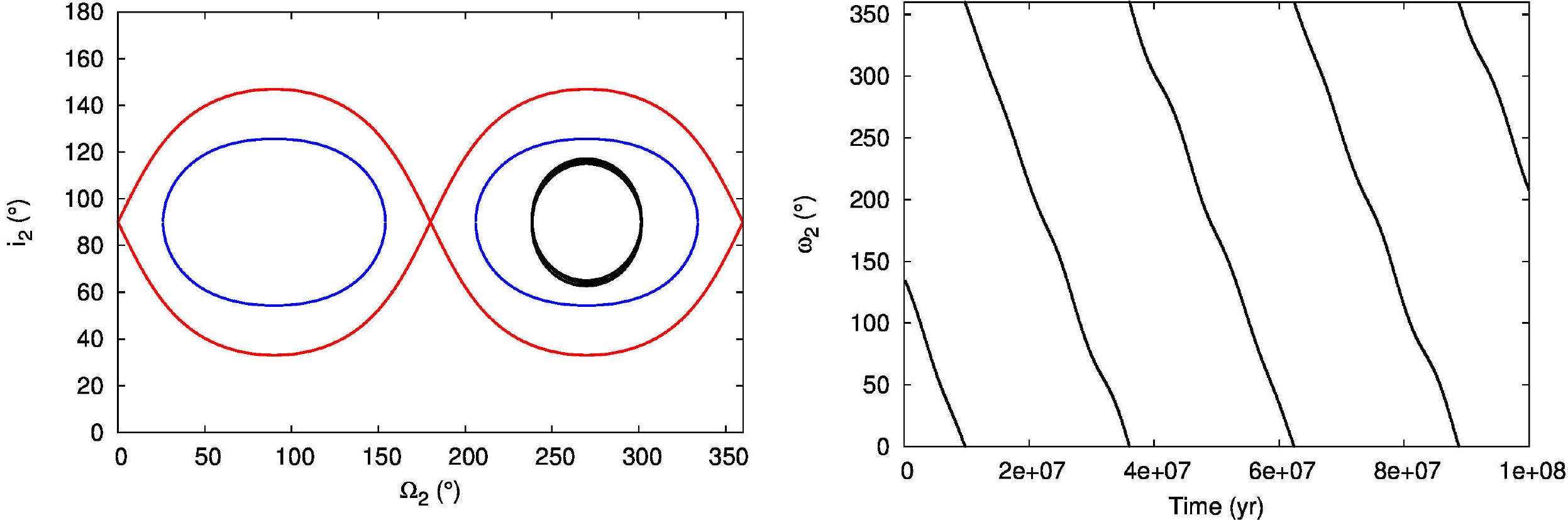}  
\caption{
{\it Left panel:} Evolutionary trajectory of a Type-F particle associated with the Set 3 of N-body simulations in the ($\Omega_2$, $i_2$) plane (black curve). The initial orbital parameters of this particle are $a_2$ = 23.580 au, $e_2$ = 0.331, $i_2$ = 72.326$^{\circ}$, $\omega_2$ = 134.249$^{\circ}$, and $\Omega_2$ = 247.646$^{\circ}$. The values of $a_1$ and $e_1$ of the inner Jupiter-mass planet of the a system are 1.333 au and 0.566, respectively. The red and blue curves illustrate the separatrix and the pairs ($\Omega_2$, $i_2$) that vanish the $\omega_2$ quadrupole precession rate for an inner perturber of $e_1$ = 0.566, respectively. {\it Right panel:} Temporal evolution of the argument of pericenter $\omega_2$ of such a particle.
}
\label{fig:fig247}
\end{figure*} 

As we said in the beginning of the present section, the inner massive perturber has an eccentricity $e_1$ greater than $e_{\text{1,F}} =$ 0.40825 in 9 of 12 N-body simulations. We also analyze in detail the evolution of the argument of pericenter $\omega_2$ of the Type-P, -R and -F particles associated with the outer small body populations of those resulting systems. 

In general terms, the evolution of the argument of pericenter $\omega_2$ of the Type-P and -R particles associated with N-body simulations of the Set 3 is very simple. In fact, Fig.~\ref{fig:fig298} illustrates the evolution of a given Type-P particle associated with a system with an inner Jupiter-mass planet whose eccentricity $e_1$ is 0.4753. In particular, the left panel exposes the trajectory of such a particle in the ($\Omega_2$, $i_2$) plane, which is represented by the black curve. The evolution of such a particle is very similar to that described for the Type-P particle illustrated in Fig.~\ref{fig:fig59}. In fact, such as the reader can see, the evolutionary trajectory of the particle never crosses the blue curve, which indicates that the $\omega_2$ quadrupole precession rate does not vanish. Finally, the right panel of Fig.~\ref{fig:fig298} shows that the argument of pericenter $\omega_2$ circulates.

It is worth mentioning that in the 9 systems under consideration with an inner massive perturber whose orbit is more eccentric than $e_{\text{1,F}} =$ 0.40825, we observe three different kinds of Type-F particles in the resulting outer reservoirs. We remark that all Type-F particles produced in our simulations show an orbit-flipping resonance that involves librations of the ascending node longitude $\Omega_2$ and variations of the inclination $i_2$ between prograde and retrograde values. The differences observed in those three kinds of Type-F particles are primarily associated with the temporal evolution of the argument of pericenter $\omega_2$.        

Figure~\ref{fig:fig247} shows the evolution of a Type-F particle representative of the first kind. In particular, the black curve in the left panel illustrates the trajectory of such a particle in the inclination $i_2$ vs. ascending node longitude $\Omega_2$ plane. While the evolution of the particle is confined to the inner region to the separatrix, it never crosses the blue curve and the $\omega_2$ quadrupole precession rate is not vanished for any pair ($\Omega_2$, $i_2$) along the entire evolution. The right panel of Fig.~\ref{fig:fig247} shows that the argument of pericenter $\omega_2$ of such a Type-F particle evolves in a circulatory regimen. 

\begin{figure}[!]
\centering
\includegraphics[width=0.48\textwidth]{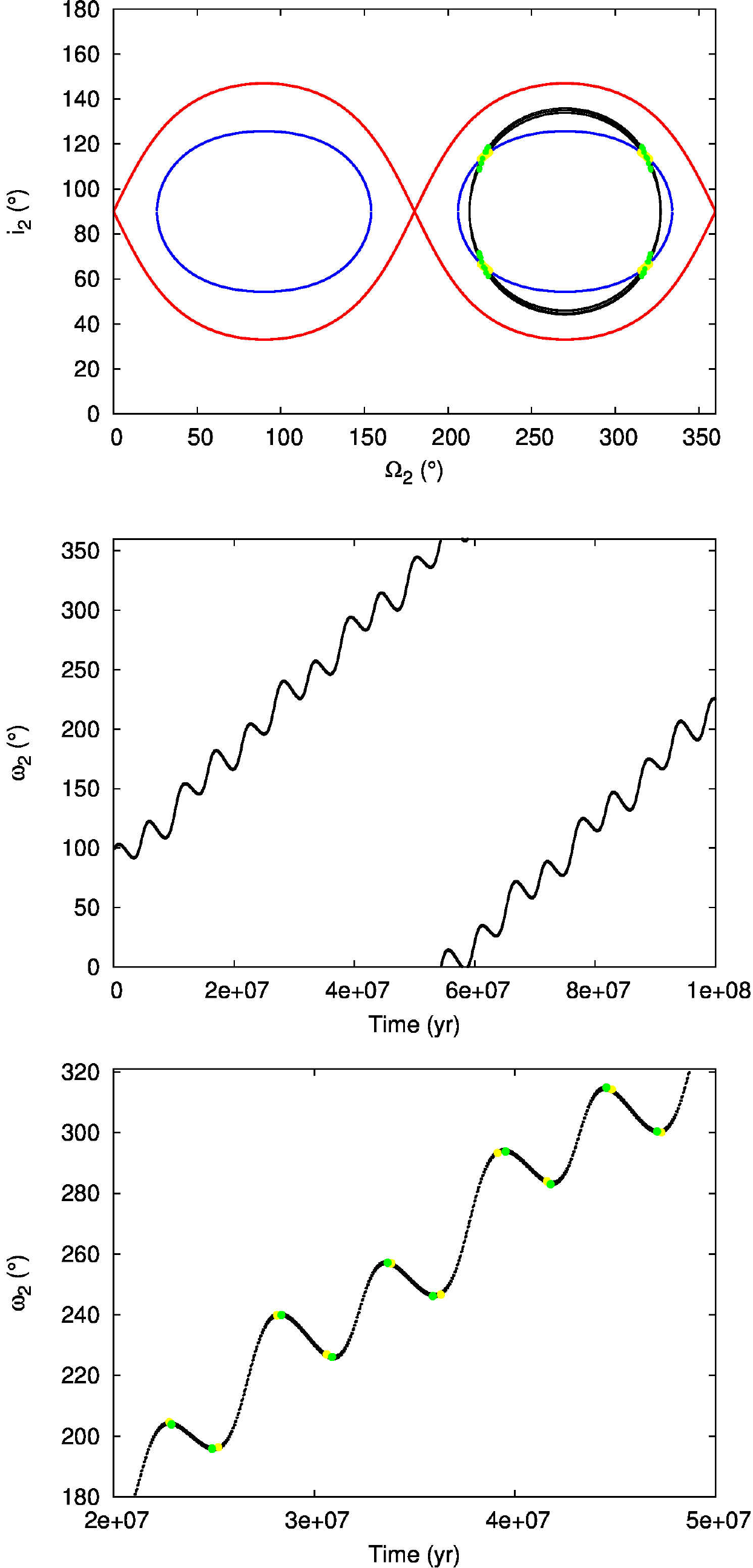} 
\caption{
{\it Top panel:} Trajectory of a Type-F particle associated with the Set 3 of N-body simulations in the ($\Omega_2$, $i_2$) plane (black curve). The initial orbital parameters of this particle are $a_2$ = 20.332 au, $e_2$ = 0.468, $i_2$ = 56.236$^{\circ}$, $\omega_2$ = 99.392$^{\circ}$, and $\Omega_2$ = 233.557$^{\circ}$. The values of $a_1$ and $e_1$ associated with the inner perturber, and the red and blue curves are defined in the caption of Fig.~\ref{fig:fig247}. {\it Middle panel:} Evolution in time of the argument of pericenter $\omega_2$ of such a particle. {\it Bottom panel:} Zoom of the middle panel between 20 Myr and 50 Myr. The references associated with the yellow and green circles illustrated in the top and bottom panels are described in the caption of Fig.~\ref{fig:fig19}. 
}
\label{fig:fig67}
\end{figure}

One representative example of a second kind of Type-F particles can be observed in Fig.~\ref{fig:fig67}. In particular, the top panel of Fig.~\ref{fig:fig67} describes the evolution of such a particle in the inclination $i_2$ vs. ascending node longitude $\Omega_2$ plane by a black curve. As the reader can see, the evolutionary trajectory periodically crosses the blue curve, which defines values associated with $i_2$ and $\Omega_2$ that vanish the $\omega_2$ quadrupole precession rate. For this particular case, the argument of pericenter $\omega_2$ circulates reaching local minimum and maximum values, which is illustrated in the middle panel of Fig.~\ref{fig:fig67}. As the reader can see, the yellow and green circles, which are associated with the vanishing of $(d\omega_{2}/d\tau)_{\text{quad}}$ (Eq.~\ref{tasaquad}) and $(d\omega_{2}/d\tau)$ (Eq.~\ref{tasatotal}), respectively, are close and they are in a good agreement with the local minimum and maximum values of $\omega_2$ observed in the bottom panel of Fig.~\ref{fig:fig67}. Moreover, the yellow and green circles are also very close in the ($\Omega_2$, $i_2$) plane, which can be seen in the top panel Fig.~\ref{fig:fig67}.

\begin{figure}[!]
\centering
\includegraphics[width=0.48\textwidth]{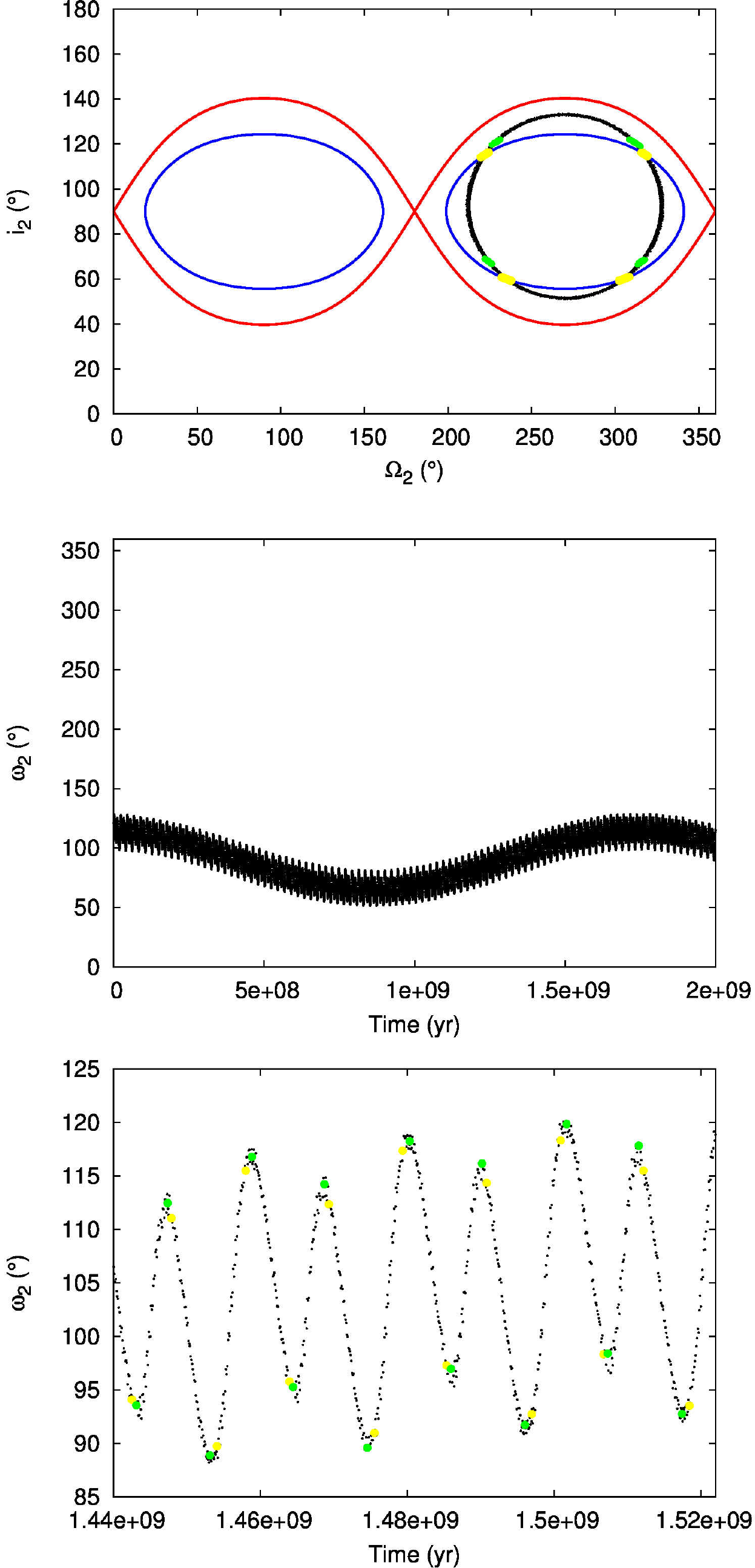} 
\caption{
{\it Top panel:} Evolutionary trajectory of a Type-F particle associated with the Set 3 of N-body simulations in the ($\Omega_2$, $i_2$) plane (black curve). The initial orbital parameters of this particle are $a_2$ = 34.201 au, $e_2$ = 0.696, $i_2$ = 52.853$^{\circ}$, $\omega_2$ = 120.147$^{\circ}$, and $\Omega_2$ = 254.671$^{\circ}$. The values of $a_1$ and $e_1$ associated with the inner perturber, and the red and blue curves are defined in the caption of Fig.~\ref{fig:fig298}. {\it Middle panel:} Evolution in time of the argument of pericenter $\omega_2$ of such a particle, which librates around 90$^{\circ}$ by defining an inverse Lidov-Kozai resonance. {\it Bottom panel:} Zoom of the middle panel between 1.44 Gyr and 1.52 Gyr. The references associated with the yellow and green circles represented in the top and bottom panels are described in the caption of Fig.~\ref{fig:fig19}.
}
\label{fig:fig914}
\end{figure}

Figure~\ref{fig:fig914} shows the evolution of a Type-F particle representative of the third kind, which is the most interesting one concerning the behavior of the argument of pericenter $\omega_2$. In particular, the evolution of such a Type-F particle in the ($\Omega_2$, $i_2$) plane is illustrated by a black curve in the top panel of Fig.~\ref{fig:fig914}. As in the previous example, the blue curve is periodically crossed by the trajectory associated with the particle, which determines pairs ($\Omega_2$, $i_2$) that lead to the vanishing of the $\omega_2$ quadrupole precession rate. However, in this particular case, the argument of pericenter $\omega_2$ librates around 90$^{\circ}$ reaching local minimum and maximum values, which can be observed in the temporal evolution of $\omega_2$ represented in the middle panel of Fig.~\ref{fig:fig914}. From this, the Type-F particles corresponding to this third kind experience an inverse Lidov-Kozai resonance associated with librations of $\omega_2$. It is important to remark that these particles show a significant interest since both the ascending node longitude $\Omega_2$ as well as the argument of pericenter $\omega_2$ evolve in a libration regime, while the inclination $i_2$ oscillates from prograde to retrograde values and back again around 90$^{\circ}$.

As in the examples represented in the Figs.~\ref{fig:fig19}, \ref{fig:fig778}, and \ref{fig:fig67}, we determine the values of $i_2$, $\Omega_2$, and $\omega_2$ of the particle's trajectory associated with the vanishing of $(d\omega_{2}/d\tau)_{\text{quad}}$ (yellow circles) and $(d\omega_{2}/d\tau)$ (green circles). On the one hand, according to that observed in the top panel of Fig.~\ref{fig:fig914}, the green and yellow circles are close but they slightly differ from each other in comparison with that observed in the top panel of Fig.~\ref{fig:fig67}, which represents the case of a Type-F particle whose argument of pericenter $\omega_2$ circulates. On the other hand, the bottom panel of Fig.~\ref{fig:fig914} shows that the local minimum and maximum values of the temporal evolution of $\omega_2$ of a Type-F particle that experiences an inverse Lidov-Kozai resonance are in a good agreement with the green circles, while the yellow ones are close.    

It is important to describe a peculiar property of the Type-F particles that experience an inverse Lidov-Kozai resonance in comparison with those Type-F particles whose argument of pericenter $\omega_2$ circulates. In fact, if $\omega_2$ evolves on a circulatory regime, the ascending node longitude $\Omega_2$ of the test particle adopts extreme values for orbital inclinations $i_2$ around 90$^{\circ}$. Thus, the evolutionary trajectory of such a particle in an inclination $i_2$ vs. ascending node longitude $\Omega_2$ plane is symmetrical respect to $i_2 =$ 90$^{\circ}$. The behavior is very different if the Type-F particle experiences an inverse Lidov-Kozai resonance. In fact, if $\omega_2$ evolves on a librating regime, the extreme values of the ascending node longitude $\Omega_2$ are not obtained for orbital inclinations $i_2$ around 90$^{\circ}$. In such a case, $\Omega_2$ adopts extreme values for prograde or retrograde inclinations depending on the center of libration associated with $\Omega_2$ and $\omega_2$. Thus, the evolutionary trajectory of such a particle in an inclination $i_2$ vs. ascending node longitude $\Omega_2$ plane evidences an asymmetry respect to $i_2 =$ 90$^{\circ}$.  

The topic concerning the asymmetrical orbital flips in a ($\Omega_2$, $i_2$) plane respect to $i_2 =$ 90$^{\circ}$ is very interesting. In fact, \citet{Zanardi2018} showed that, when the general relativity is included, the extreme values of the ascending node longitude $\Omega_2$ associated with a given outer test particle are obtained for retrograde inclinations, leading to an asymmetrical orbital flip in a ($\Omega_2$, $i_2$) plane respect to $i_2 =$ 90$^{\circ}$. From this, it is very important to remark that the results derived in the present research indicate that, in absence of general relativity, the existence of an asymmetrical orbital flip in a ($\Omega_2$, $i_2$) plane respect to an inclination $i_2 =$ 90$^{\circ}$ is a clear sign of an inverse Lidov-Kozai resonance.     

We can understand the correlation between the asymmetrical orbital flips and the inverse Lidov-Kozai resonance from analytical considerations. To do this, it is important to remark that it is not enough to analyze the extreme values of the ascending node longitude $\Omega_2$ from the vanishing of the $\Omega_2$ quadrupole precession rate. In fact, such as we mention in Sect. 2, the $\Omega_2$ quadrupole precession rate vanishes for an inclination $i_2 =$ 90$^{\circ}$, which would lead to symmetrical orbital flips in a ($\Omega_2$, $i_2$) plane respect to $i_2 =$ 90$^{\circ}$, regardless of the evolution of the argument of pericenter $\omega_2$. To describe in detail the symmetry or asymmetry of the trajectories associated with Type-F particles in an inclination $i_2$ vs. ascending node longitude $\Omega_2$ plane respect to $i_2 =$ 90$^{\circ}$, we must compute the extreme values of the ascending node longitude $\Omega_2$ from the vanishing of the $\Omega_2$ precession rate calculated (at least) up to the octupole level of the secular approximation. From Eq. 4 and the expression of the energy function given by Eq.~\ref{fhastaoct}, the $\Omega_2$ precession rate computed up to the octupole level of the approximation is expressed by
\begin{eqnarray}
\frac{d\Omega_2}{d\tau} &=& -\frac{\partial f_{\text{quad}}}{\partial \theta_2}\frac{1}{J_2}-\frac{\partial f_{\text{oct}}}{\partial \theta_2}\frac{\epsilon}{J_2} \nonumber \\
&=& \frac{1}{(1-e^2_2)^2}\Bigg\{\alpha\theta_2^2 + \beta\theta_2 + \gamma\Bigg\},  
\label{tasanodo}
\end{eqnarray}
where
\begin{eqnarray}
\alpha &=& \frac{15}{4}e_1\epsilon\Bigg[-45-\frac{15}{2}e^2_1+\frac{105}{2}e^2_1\cos(2\Omega_2)\Bigg]\sin\omega_2\sin\Omega_2,  \nonumber \\
\beta &=& -6(2+3e^2_1)+30e^2_1\cos(2\Omega_2) \nonumber \\
&+&\frac{15}{4}e_1\epsilon\Bigg[5(2+5e^2_1)-35e^2_1\cos(2\Omega_2)\Bigg]\cos\omega_2\cos\Omega_2, \nonumber \\
\gamma &=& \frac{15}{4}e_1\epsilon\Bigg[11-\frac{e^2_1}{2}-\frac{35}{2}e^2_1\cos(2\Omega_2)\Bigg] \sin\omega_2\sin\Omega_2. \nonumber \\ 
\label{coeficientestasanodo}
\end{eqnarray}
For the particular case of Type-F particles, the ascending node longitude $\Omega_2$ evolves in a librating regime. The presence of sines and cosines of the argument of pericenter $\omega_2$ in $\alpha$, $\beta$, and $\gamma$ parameters indicates that the vanishing of the $\Omega_2$ precession rate depends on the evolutionary regime associated with $\omega_2$. 

\begin{figure*}[!]
\centering
\includegraphics[width=0.95\textwidth]{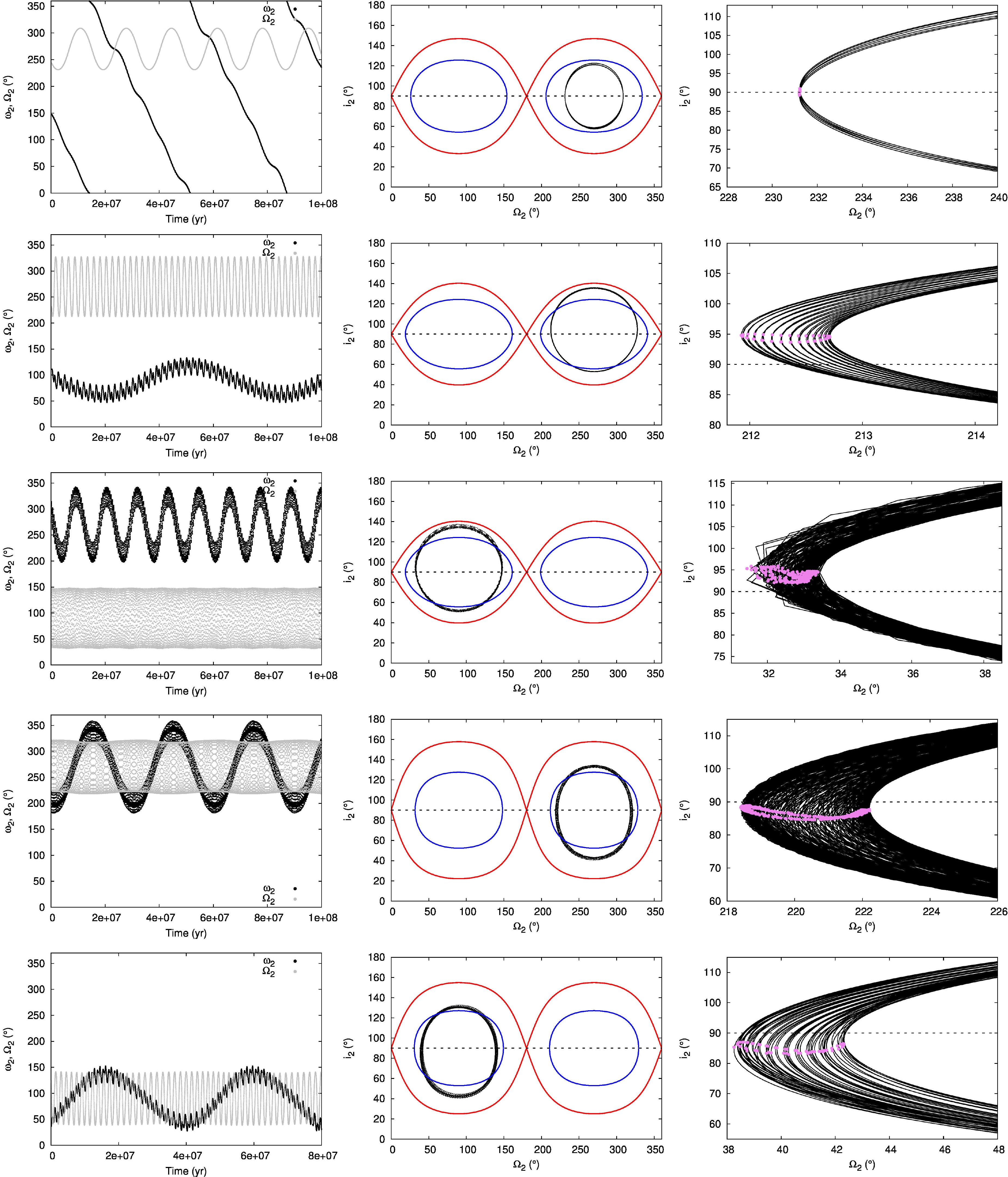} 
\caption{
Evolution of different Type-F particles associated with the Set 3 of N-body simulations. The rows are numerated from 1 to 5 from top to bottom. In every row, the black and green curves of the left panel illustrate the evolution in time of $\omega_2$ and $\Omega_2$, respectively. Then, the black and red curves of the middle panel represent the trajectory of the particle and the separatrix in the plane ($\Omega_2$, $i_2$), respectively, while the blue curve shows the pairs ($\Omega_2$, $i_2$) of the particle's trajectory that vanish the $\omega_2$ quadrupole precession rate of the system under consideration. Finally, the violet circles of the right panel illustrate the pairs ($\Omega_2$, $i_2$) of the particle's trajectory that vanish the $\Omega_2$ precession rate calculated up to the octupole level of the approximation. The initial orbital parameters of the test particles and the values of $a_1$ and $e_1$ associated with the inner Jupiter-mass planet are specified for every row in Table~\ref{TablaParametros}.     
}
\label{fig:figcombo}
\end{figure*}

To understand this, we analyze 5 Type-F particles of the Set 3 of N-body simulations with different behaviors concerning the evolution of the argument of pericenter $\omega_2$, which are illustrated in Fig.~\ref{fig:figcombo}. This figure shows 5 rows, which are numbered from 1 to 5 from top to bottom. Every row refers to a given Type-F particle. The left panels of each row describe the evolution in time of the argument of pericenter $\omega_2$ and the ascending node longitude $\Omega_2$ of the test particle under study by black and gray curves, respectively. Then, the middle panels represent the evolutionary trajectory in the ($\Omega_2$, $i_2$) plane of each Type-F particle by a black curve, while the separatrix and the pairs ($\Omega_2$, $i_2$) that vanish the $\omega_2$ quadrupole precession rate for the system under consideration are also illustrated by red and blue curves, respectively. Finally, the right panels display a zoom of the middle panels by including the pairs ($\Omega_2$, $i_2$) of the trajectory of the corresponding test particle that vanish the $\Omega_2$ precession rate computed up to the octupole level of the approximation (Eq.~\ref{tasanodo}), which are represented by violet circles. Table~\ref{TablaParametros} summarizes the initial orbital parameters associated with the 5 Type-F particles analyzed in Fig.~\ref{fig:figcombo}, and the values of the semimajor axis $a_1$ and the eccentricity $e_1$ of the inner Jupiter-mass planet of each system corresponding to such particles.

The row 1 of Fig.~\ref{fig:figcombo} illustrates the particular case of a Type-F particle whose argument of pericenter $\omega_2$ circulates. According to this, the extreme values of the ascending node longitude $\Omega_2$ are obtained for inclinations $i_2$ around 90$^{\circ}$, which leads to symmetrical orbital flips in the ($\Omega_2$, $i_2$) plane respect to $i_2 =$ 90$^{\circ}$. 

The row 2 of Fig.~\ref{fig:figcombo} describes the evolution of a Type-F particle whose $\omega_2$ ($\Omega_2$) librates around 90$^{\circ}$ (270$^{\circ}$). From this, the extreme values of $\Omega_2$ are associated with retrograde inclinations, which produces asymmetrical orbital flips in the ($\Omega_2$, $i_2$) plane respect to $i_2 =$ 90$^{\circ}$. This also can be observed in the row 3 of Fig.~\ref{fig:figcombo}. In this case, the particle of study experiences librations of $\omega_2$ ($\Omega_2$) around 270$^{\circ}$ (90$^{\circ}$). As in the row 2, retrograde inclinations determine the extreme values of $\Omega_2$, which leads to an asymmetry in the evolutionary trajectory of the Type-F particle in the ($\Omega_2$, $i_2$) plane respect to $i_2 =$ 90$^{\circ}$.         

The rows 4 and 5 of Fig.~\ref{fig:figcombo} show a different result in comparison with that above described. In fact, both the argument of pericenter $\omega_2$ as well as the ascending node longitude $\Omega_2$ of the Type-F particle analyzed in the row 4 (5) librate around 270$^{\circ}$ (90$^{\circ}$). When $\omega_2$ and $\Omega_2$ have the same center of libration, the extreme values of $\Omega_2$ are associated with prograde inclinations, which produces asymmetrical orbital flips in the ($\Omega_2$, $i_2$) plane respect to $i_2 =$ 90$^{\circ}$.     

We would like to remark that the test particle represented in the row 4 of Fig.~\ref{fig:figcombo} is associated with a system with an inner massive perturber whose eccentricity $e_1$ is of 0.739. This simple example is of significant interest and it illustrates that the inverse Lidov-Kozai resonance can be found even for high values of the inner perturber's eccentricity $e_1$ from a suitable selection of initial conditions.

\section{Discussion and conclusions}

In the present research, we study the evolution of the argument of pericenter $\omega_2$ of outer test particles that orbit a given central star and undergo the effects of an inner massive perturber. In particular, we describe the behavior of $\omega_2$ as a function of the orbital eccentricity $e_1$ of the inner perturber. The key result derived in our study indicates that the inverse Lidov-Kozai or $\omega_2$ resonance can appear for small, moderate, and high values of $e_1$ as long as suitable initial conditions mainly associated with $i_2$ and $\Omega_2$ are adopted. 

First, we carry out our investigation from analytical considerations. To do this, we adopt the expression of the potential expanded up to the octupole level of the secular approximation derived by \citet{Naoz2017}, as well as the term of hexadecapolar order included later by \citet{Vinson2018}. From this, we compute equations that express the contribution of the terms of quadrupole, octupole, and hexadecapolar order of the secular approximation to the $\omega_2$ precession rate. Our study suggests that the pairs ($\Omega_2$, $i_2$) that vanish the $\omega_2$ quadrupole precession rate $(d\omega_2/d\tau)_{\text{quad}}$ strongly depend on the eccentricity $e_1$ of the inner perturber. In fact, if $e_1 < 0.25$, $(d\omega_2/d\tau)_{\text{quad}}$ is only vanished for test particles on prograde and retrograde orbits whose ascending node longitude $\Omega_2$ evolves in a circulatory regime, while, if $e_1 > 0.40825$, $(d\omega_2/d\tau)_{\text{quad}}$ is only vanished for test particles that experience an orbit-flipping resonance, in which $\Omega_2$ librates. For inner perturber eccentricities $e_1$ between 0.25 and 0.40825, our analysis indicates that any test particle can vanish the $\omega_2$ quadrupole precession rate regardless the evolutionary regime of the ascending node longitude $\Omega_2$, for pairs ($\Omega_2$, $i_2$) that satisfy the relation given by Eq.~\ref{anulacionquadi}.

Furthermore, we use the analytical considerations derived in our research with the aim of describing the behavior of test particles, which result from a set of N-body simulations presented by \citet{Zanardi2017}. On the basis of such simulations, we analyze the evolution of the argument of pericenter $\omega_2$ of outer test particles that evolve under the effects of a Jupiter-mass planet around a 0.5 M$_{\odot}$ star. The eccentricity $e_1$ of the inner perturber associated with the sample of simulations of work ranges from 0.227 to 0.94. We remark that the evolution of $\omega_2$ of the outer test particles extracted from the N-body simulations carried out \citet{Zanardi2017} are in a very good agreement with the analytical criteria derived in the present investigation. 

It is very important to mention that, unlike that proposed by \citet{Vinson2018}, who found the inverse Lidov-Kozai resonance only up to $e_1 =$ 0.1 in N-body experiments, we observe outer test particles that experience an $\omega_2$ resonance for an inner perturber's eccentricity $e_1$ as high as 0.8 in the N-body simulations that represent our frame of work. As we remarked in the beginning of this section, our research indicates that the inverse Lidov-Kozai resonance can be found even for high values of $e_1$ as long as a correct choice of the initial conditions is made.

\begin{table*}
\caption{Initial orbital parameters concerning the semimajor axis $a_2$, eccentricity $e_2$, inclination $i_2$, argument of pericenter $\omega_2$, and ascending node longitude $\Omega_2$ associated with the Type-F particles of Fig.~\ref{fig:figcombo}. Moreover, the semimajor axis $a_1$ and the eccentricity $e_1$ of the inner Jupiter-mass planet of each system are also specified. The rows are numerated from 1 to 5 from top to bottom of Fig.~\ref{fig:figcombo}. It is important to remark that the Type-F particles corresponding to the rows 2, 3, 4, and 5 show values associated with $(a_1/a_2)$ or/and $\epsilon$ parameter slightly greater than 0.1. 
}
\begin{center}
\begin{tabular}{cccccccc}
\hline
\hline \\
Row  & $a_1$ (au) & $e_1$ & $a_2$ (au) & $e_2$ & $i_2$ ($^{\circ}$) &  $\omega_2$ ($^{\circ}$) &   $\Omega_2$ ($^{\circ}$)     \\
\hline
\hline \\
1 & 1.333 & 0.566 & 22.812 & 0.310 & 66.702 & 147.045 & 244.081   \\
2 & 1.778 & 0.475 & 17.374 & 0.676 & 64.417 & 110.960 & 231.042   \\
3 & 1.778 & 0.475 & 9.017 & 0.458 & 65.22 & 274.070 & 130.580   \\
4 & 0.977 & 0.739 & 7.127 & 0.513 & 132.016 & 218.093 & 288.414   \\
5 & 1.487 & 0.691 & 19.102 & 0.700 & 51.499 & 36.313 & 124.091   \\
\hline
\hline
\end{tabular}
\end{center}
\label{TablaParametros}
\end{table*}

It is worth noting that the inverse Lidov-Kozai resonance produces some distinctive features in the evolution of a test particle in the inclination $i_2$ .vs. ascending node longitude $\Omega_2$ plane. On the one hand, if a given particle experiences an inverse Lidov-Kozai resonance and its ascending node longitude $\Omega_2$ evolves in a circulatory regime, the extreme values of the inclination $i_2$ at $\Omega_2 =$ 90$^{\circ}$ and 270$^{\circ}$ are not equal. Our study shows that the value of $\Omega_2$ associated with the absolute extreme of $i_2$ depends on the center of libration of $\omega_2$. On the other hand, if a test particle is in an inverse Lidov-Kozai resonance and its ascending node longitude $\Omega_2$ librates, the evolutionary trajectory of such a particle in the inclination $i_2$ .vs. ascending node longitude $\Omega_2$ plane evidences an asymmetry respect to $i_2 =$ 90$^{\circ}$. Our analysis shows that the extreme values of $\Omega_2$ are obtained for values of $i_2$ less or higher than 90$^{\circ}$, which depends on the centers of libration associated with $\Omega_2$ and $\omega_2$. We remark that such distinctive features observed in the evolution of test particles extracted from the N-body experiments are very good described from the analytical expressions derived in our investigation.

It is important to mention that our investigation shows that the vanishing of $\omega_2$ quadrupole precession rate is not a sufficient condition for the inverse Lidov-Kozai resonance. A more detailed study is beyond the scope of this paper. 

The dynamics discussed in this research could play a key role in understanding the evolution of debris disks associated with extrasolar systems that host an inner and eccentric giant planet. In this sense, planetary systems such as HD 10647, HD 39091, HD 50499, HD 50554, and HD 210277 can serve as valuable laboratories to contrast theoretical results with observational evidence from dynamical and collisional models associated with the evolution of debris disks.

The present work represents a detailed investigation that combines analytical considerations and numerical results derived from N-body simulations concerning the inverse Lidov-Kozai resonance. Such a treatment allows us obtain a better understanding about the evolution of the argument of pericenter $\omega_2$ of an outer test particle in the elliptical restricted three-body problem. 

\begin{acknowledgements}
This work was partially financed by CONICET through PIP 0436/13, and Agencia de Promoci\'on Cient\'{\i}fica, through PICT 2014-1292 and PICT 201-0505. We thank the anonymous referee for valuable suggestions, which helped us to improve the manuscript. Moreover, G.C.dE, M.Z., and A.D. acknowledge the financial support by FCAGLP and IALP for extensive use of their computing facilities, and S.N. acknowledges the partial support from the NSF through Grant No. AST-173916. Finally, G.C.dE and M.Z. wish to dedicate the present paper to the memory of Carlos Rafael de El\'{\i}a.
\end{acknowledgements}

%%%%%%%%%%%%%%%%%%%%%%%%%%%%%%%%%%%%%%%%%%%%%%%%%%%%%%%%%%%%
% BIBLIOGRAFIA
%%%%%%%%%%%%%%%%%%%%%%%%%%%%%%%%%%%%%%%%%%%%%%%%%%%%%%%%%%%%
\bibliographystyle{aa}        % style aa.bst
\bibliography{paper} % your references Yourfile.bib

\end{document}